\documentclass[11pt]{article}
\usepackage[T1]{fontenc}      
\usepackage{latexsym}
\usepackage{amssymb}
\usepackage{amsmath}
\usepackage{graphicx}
\usepackage{colortbl}
\usepackage{tikz}

\usepackage[square,comma]{natbib} 

\begin{document}

\def\Nset{\mathbb{N}}
\def\Ascr{\mathcal{A}}
\def\Bscr{\mathcal{B}}
\def\Cscr{\mathcal{C}}
\def\Dscr{\mathcal{D}}
\def\Escr{\mathcal{E}}
\def\Fscr{\mathcal{F}}
\def\Hscr{\mathcal{H}}
\def\Iscr{\mathcal{I}}
\def\Mscr{\mathcal{M}}
\def\Nscr{\mathcal{N}}
\def\Pscr{\mathcal{P}}
\def\Cscr{\mathcal{C}}
\def\Rscr{\mathcal{R}}
\def\Sscr{\mathcal{S}}
\def\Uscr{\mathcal{U}}
\def\Wscr{\mathcal{W}}
\def\Xscr{\mathcal{X}}
\def\cupp{\stackrel{.}{\cup}}
\def\stern{\textasteriskcentered}
\def\bold{\bf\boldmath}
\def\tp{{\scriptscriptstyle\top}}
\def\Ssumeven{\sum_{S\in\Sscr \;\! : \;\! |S\cap C| \,\text{even}}}
\def\Ssumone{\sum_{S\in\Sscr \;\! : \;\! |S\cap C|=1}}
\def\Csumeven{\sum_{C\in\Cscr \;\! : \;\! |S\cap C| \,\text{even}}}
\def\Csumone{\sum_{C\in\Cscr \;\! : \;\! |S\cap C|=1}}
\def\bomc{\text{\rm BOMC}}

\newcommand{\todo}[1]{\marginpar{\textcolor{red}{\small #1}}}
\newcommand{\rouge}[1]{\textcolor{red}{\tt \footnotesize #1}}

\newcommand{\boldheader}[1]{\smallskip\noindent{\bold #1:}\quad}
\newcommand{\PP}{\mbox{\slshape P}}
\newcommand{\NP}{\mbox{\slshape NP}}
\newcommand{\opt}{\mbox{\scriptsize\rm OPT}}
\newcommand{\ec}{\mbox{\scriptsize\rm OPT}_{\small\rm 2EC}}
\newcommand{\lp}{\mbox{\scriptsize\rm LP}}
\newcommand{\inn}{\mbox{\rm in}}
\newcommand{\deff}{\mbox{\rm sur}}
\newcommand{\MAXSNP}{\mbox{\slshape MAXSNP}}
\newtheorem{theorem}{Theorem}
\newtheorem{lemma}[theorem]{Lemma}
\newtheorem{corollary}[theorem]{Corollary}
\newtheorem{proposition}[theorem]{Proposition}
\newtheorem{definition}[theorem]{Definition}
\def\prove{\par \noindent \hbox{\bf Proof:}\quad}
\def\endproof{\eol \rightline{$\Box$} \par}
\renewcommand{\endproof}{\hspace*{\fill} {\boldmath $\Box$} \par \vskip0.5em}
\newcommand{\mathendproof}{\vskip-1.8em\hspace*{\fill} {\boldmath $\Box$} \par \vskip1.8em}
\def\cupp{\stackrel{.}{\cup}}

\newcommand{\citegenitiv}[1]{\citeauthor{#1}' [\citeyear{#1}]}
\newcommand{\citegenitivs}[1]{\citeauthor{#1}'s [\citeyear{#1}]}

\definecolor{orange}{rgb}{1,0.5,0}
\definecolor{violet}{rgb}{0.8,0,1}
\definecolor{darkgreen}{rgb}{0,0.5,0}
\definecolor{grey}{rgb}{0.6,0.6,0.6}
\definecolor{lightgrey}{rgb}{0.7,0.7,0.7}
\definecolor{turq}{rgb}{0,0.4,0.4}

\setcounter{topnumber}{9}
\setcounter{bottomnumber}{9}
\setcounter{totalnumber}{9}
\renewcommand{\topfraction}{0.99}
\renewcommand{\bottomfraction}{0.99}
\renewcommand{\textfraction}{0.01}

\title {
\vspace*{-1.8cm}
{\bf\boldmath Reassembling Trees for the Traveling Salesman} 
}
\author{\Large Jens Vygen \\[1mm] \small University of Bonn}
\date{\small revised version -- December 23, 2015}

\begingroup
\makeatletter
\let\@fnsymbol\@arabic
\maketitle
\endgroup

\begin{abstract}
Many recent approximation algorithms for different variants of the traveling salesman problem 
(asymmetric TSP, graph TSP, $s$-$t$-path TSP) exploit the well-known fact that a solution of the natural
linear programming relaxation can be written as convex combination of spanning trees.
The main argument then is that randomly sampling a tree from such a distribution and then 
completing the tree to a tour at minimum cost yields a better approximation guarantee than simply
taking a minimum cost spanning tree (as in Christofides' algorithm).

We argue that an additional step can help: reassembling the spanning trees before sampling.
Exchanging two edges in a pair of spanning trees can improve their properties under certain conditions.

We demonstrate the usefulness for the metric $s$-$t$-path TSP by devising a deterministic polynomial-time
algorithm that improves on Seb\H{o}'s previously best approximation ratio of $\frac{8}{5}$.

\medskip\noindent
\noindent{{\bf keywords:} traveling salesman problem, $s$-$t$-path TSP, approximation algorithm, spanning tree}
\end{abstract}

\section{Introduction}

The traveling salesman problem (TSP) is probably the best-known $\NP$-hard combinatorial optimization problem.
Although for the general metric TSP \citegenitiv{Chr76} algorithm with its approximation ratio $\frac{3}{2}$ is
still unbeaten, we have seen progress for several variants and special cases in particular since 2010.
See \cite{Vyg12} for a detailed survey.

Many of the recent approximation algorithms begin by 
solving the natural linear programming relaxation, which was first proposed by \cite{DanFJ54}.
It was observed by \cite{HelK70} that a solution $x^*$ to this LP can (after scaling down by a factor $\frac{n-1}{n}$ except for the
$s$-$t$-path case) be written as convex combination (or, equivalently, probability distribution) of spanning trees. 
Of course, this distribution is far from unique. 

\cite{AsaXX10} and \cite{OveSS11} improved the approximation ratio for the asymmetric TSP and graph TSP, respectively,
by randomly sampling a spanning tree from a maximum entropy distribution describing $\frac{n-1}{n}x^*$ and then completing it to
a tour in an optimal way. 
\cite{AnKS12} considered the metric $s$-$t$-path TSP and showed that 
a spanning tree randomly chosen from \emph{any} distribution describing $x^*$ is ---in expectation--- good
enough to improve on Christofides' algorithm for this problem. 

In this paper, we propose to modify the distribution before sampling.
By exchanging two edges in a pair of trees with certain properties we obtain two new trees and hence a new distribution.
We call this step \emph{reassembling trees}. 
Under certain conditions the two new trees have better properties than the old ones.
For the $s$-$t$-path TSP we show that this step can indeed improve the approximation ratio.

\subsection{The \boldmath{$s$-$t$}-path TSP}

Let us define the (metric) $s$-$t$-path TSP formally.
As in the classical version of the TSP, we want to visit a set of cities at minimum total cost.
However, rather than returning to the origin at the end, we are given the origin and the destination as input.
More precisely,
we are given a finite set $V$, two elements $s$ and $t$ of $V$, and a symmetric distance function
$c:V\times V\to\mathbb{R}_{\ge 0}$ satisfying the triangle inequality.
Throughout this paper we will denote by $n$ the number of elements of $V$.
We ask for a sequence $V=\{v_1,\ldots,v_n\}$ with $v_1=s$ and $v_n=t$, such that
$\sum_{i=1}^{n-1} c(v_i,v_{i+1})$ is minimized.

The classical metric TSP is the special case when $s=t$.
We note that in both variants we can either require that every city is visited exactly once, or, 
equivalently, we can drop the requirement that $c$ satisfies the triangle inequality but allow visiting cities more than once. 

The $s$-$t$-path TSP (for $s\not=t$) is clearly no easier than the classical metric TSP (we can reduce the latter to the former
by guessing two cities that are adjacent in an optimum tour); in particular there is no
polynomial-time approximation algorithm with better ratio than $\frac{123}{122}$ unless $\PP=\NP$
(\cite{KarLS13}).

We assume $s\not=t$ henceforth.
It is sufficient to compute a connected multi-graph with vertex set $V$ in which exacty $s$ and $t$ have odd degree;
such a graph will be called an \emph{$\{s,t\}$-tour}. Here is why:
given an $\{s,t\}$-tour, we can find an Eulerian walk from $s$ to $t$ (using every edge exactly once) in linear time
and shortcut whenever a vertex is visited not for the first time;
this yields an $s$-$t$-path with vertex set $V$, which ---due to the triangle inequality--- is no more expensive than 
the $\{s,t\}$-tour.

\subsection{Previous approximation algorithms}

\citegenitiv{Chr76} algorithm, originally designed for the classical metric TSP,
works also for the $s$-$t$-path TSP. It first computes a minimum cost spanning tree $(V,S)$
in the complete graph spanned by $V$, and then adds a minimum cost $T_S$-join $J$, where
$T_S$ is the set of vertices whose degree has the wrong parity (even for $s$ or $t$, odd for other vertices).
The result is an $\{s,t\}$-tour. 

\cite{Hoo91} showed that Christofides' algorithm has approximation ratio $\frac{5}{3}$ for the $s$-$t$-path TSP,
and that in fact this ratio is asymptotically attained by an infinite set of examples.

\cite{AnKS12} proposed the  \emph{best-of-many Christofides algorithm} for the $s$-$t$-path TSP
and proved that it has approximation ratio $\frac{1+\sqrt{5}}{2}\approx 1.619$.
The algorithm is quite simple:
it computes an optimum solution to the natural LP relaxation
(see (\ref{stpathlp}) below) and writes it as convex combination of spanning trees.
For each of these spanning trees, $(V,S)$, it computes a minimum weight $T_S$-join $J$, where
$T_S$ is the set of vertices whose degree has the wrong parity, 
obtains an $\{s,t\}$-tour $S\cupp J$,
and finally outputs the best of these.
We will go into details in Subsection \ref{bomc}.
This was the first improvement over Christofides' algorithm that applied to general metrics.
Then \cite{Seb13} improved the analysis, obtaining the approximation ratio $\frac{8}{5}$.
We will describe how in Subsection \ref{oldanalysis}.

For the special case where $c$ is the metric closure of an unweighted graph,
better approximation algorithms have been obtained by \cite{MomS11}, \cite{Muc12}, and \cite{AnKS12} (in this order).
The best known approximation ratio 1.5, obtained first 
by \cite{SebV12}, matches the integrality ratio of the LP in this special case.
\cite{Gao13} gave a simpler proof of this result.

Some of the above-mentioned papers apply also to a generalization (to $T$-tours, for general $T$); 
see Section \ref{conclusion} for a brief discussion.


\subsection{Notation}

For a given instance $(V,s,t,c)$ let $n:=|V|$, and let $E:={V \choose 2}$ be the edge set of the complete graph on $V$.
For $U\subseteq V$, $\delta(U)$ denotes the set of edges with exactly one endpoint in $U$,
and $E[U]$ the set of edges with both endpoints in $U$.
We write $\delta(v):=\delta(\{v\})$ for $v\in V$.
For $T\subseteq V$ with $|T|$ even, a \emph{$T$-join} is a set $J\subseteq E$ for which 
$|\delta(v)\cap J|$ is odd for all $v\in T$ and
$|\delta(v)\cap J|$ is even for all $v\in V\setminus T$. 
A \emph{$T$-cut} is a cut $\delta(U)$ for which $|U\cap T|$ is odd.
The intersection of a $T$-join and a $T$-cut always contains an odd number of edges.
\cite{Edm65} proved that a minimum weight $T$-join can be computed in polynomial time.

For a vector $x\in\mathbb{R}^E$ and $F\subseteq E$
we write $x(F):=\sum_{e\in F} x_e$
and $c(x):=\sum_{e=\{v,w\}\in E} c(v,w)x_e$;
moreover, $\chi^F\in\{0,1\}^E$ denotes the incidence vector of $F$ 
(i.e., $\chi^F_e=1$ for $e\in F$ and $\chi^F_e=0$ for $e\in E\setminus F$),
and $c(F):=c(\chi^F)=\sum_{e=\{v,w\}\in F} c(v,w)$.
By $\Sscr$ we denote the set of edge sets of spanning trees in $(V,E)$.
For $S\in\Sscr$, the set 
$T_S:=\{v\in V\setminus\{s,t\}: |\delta(v)\cap S| \text{ odd}\}\cup\{v\in \{s,t\}: |\delta(v)\cap S| \text{ even}\}$
contains the vertices whose degree in $S$ has the wrong parity.

\subsection{Best-of-many Christofides \label{bomc}}

Our algorithm will be an extension of the \emph{best-of-many Christofides algorithm}
that was proposed by \cite{AnKS12}. Therefore, we first describe this algorithm in more detail.

The algorithm begins by solving the LP relaxation 
\begin{equation}
\label{stpathlp}
\hspace*{-0.4cm}
\begin{array}{lcrclcl}
\multicolumn{2}{l}{\min \ c(x)} && & & \\[0.5mm]
\mbox{subject to} && x(\delta(U)) &\ge& 2 && (\emptyset\not=U\subset V,\, |U\cap \{s,t\}| \text{ even}) \\
&& x(\delta(U)) &\ge& 1 && (\emptyset\not=U\subset V,\, |U\cap \{s,t\}| \text{ odd}) \\
&& x(\delta(v)) &=& 2 && (v\in V\setminus \{s,t\}) \\
&& x(\delta(v)) &=& 1 && (v\in \{s,t\}) \\
&& x_e & \ge & 0 && (e\in E)
\end{array}
\hspace*{-0.4cm}
\end{equation}

Obviously, the integral solutions to (\ref{stpathlp}) are precisely the incidence vectors
of the edge sets of the Hamiltonian $s$-$t$-paths in $(V,E)$. So this LP is indeed a relaxation.

Following an idea of \cite{HelK70}, \cite{AnKS12} observed that the polytope defined by (\ref{stpathlp}) is 
$\bigl\{x\in\mathbb{R}_{\ge 0}^E: x(E)=n-1,\, 
x(E[U])\le |U|-1 \ \forall\, \emptyset\not=U\subset V,\, 
x(E[U])\le |U|-2 \ \forall\, \{s,t\}\subseteq U\subset V \bigr\}$ 
and hence is contained in the spanning tree polytope of $(V,E)$ (\cite{Edm70}).
Therefore, an optimum solution $x^*$ (in fact every feasible solution) can be written as 
$x^*=\sum_{S\in\Sscr}p_S \chi^S$,
where $p$ is a \emph{distribution} on $\Sscr$, 
i.e., $p_S\ge 0$ for all $S\in\Sscr$
and $\sum_{S\in\Sscr}p_S=1$.

By Carath\'eodory's theorem we can assume that $p_S>0$ for less than $n^2$ spanning trees $(V,S)$.  
An optimum LP solution $x^*$, such spanning trees, and such numbers $p_S$ can be computed in polynomial time,
as can be shown with the ellipsoid method (\cite{GroLS81}) or using the splitting-off technique (cf.\ \cite{GenW15}).

We will fix $x^*$ henceforth. We will also fix the distribution $p$ for the rest of the introduction but will modify it later.

For each $S\in\Sscr$ with $p_S>0$, the best-of-many Christofides algorithm then computes a minimum weight $T_S$-join $J$,
and considers
the $\{s,t\}$-tour $S\cupp J$; the output is the best of these.
Note that trying all $S\in\Sscr$ with $p_S>0$ leads to a deterministic polynomial-time algorithm 
which is at least as good as randomly picking $S\in\Sscr$ with probability $p_S$.

\subsection{Basic Analysis}

We follow the basic analysis of \cite{AnKS12}.
The cost of the $\{s,t\}$-tour that the best-of-many Christofides algorithm computes for a given instance
depends on its choice of $p$ only:
it is
$\bomc(p):=\min_{S\in\Sscr:\, p_S>0} \left(c(S)+\min\{c(J): J \mbox{ is a $T_S$-join}\}\right)$.
This is at most 
$\sum_{S\in\Sscr} p_S \left(c(S)+\min\{c(J): J \mbox{ is a $T_S$-join}\}\right)$.
Note that $\sum_{S\in\Sscr} p_S c(S) = c(x^*)$.

A well-known result of \cite{EdmJ73} says that the minimum weight of a $T_S$-join is
the minimum $c(y)$ over all $y$ in the $T_S$-join polyhedron
\begin{equation}
\label{TSjoinpolyhedron}
\left\{y\in\mathbb{R}^E_{\ge 0}: y(C)\ge 1 \ \forall \text{ $T_S$-cuts } C \right\}.
\end{equation}

Therefore
\begin{equation}
\label{basicbound}
\bomc(p) \ \le \ c(x^*) + \sum_{S\in\Sscr} p_S c(y^S)
\end{equation}
for any set of vectors $(y^S)_{S\in\Sscr}$ such that $y^S$ is in the $T_S$-join polyhedron
(\ref{TSjoinpolyhedron}).
The difficulty in the analysis lies in finding an appropriate set of vectors $(y^S)_{S\in\Sscr}$.
Let us call them \emph{correction vectors} (although \cite{AnKS12} used this term with a different meaning),
because they bound the cost of parity correction.

\subsection{Narrow cuts}

Let $\Cscr:=\{\delta(U):\emptyset\not=U\subset V,\,  x^*(\delta(U))<2\}$.
The elements of $\Cscr$ are called \emph{narrow cuts}.
By (\ref{stpathlp}) they are $\{s,t\}$-cuts.
If there are no narrow cuts except $\delta(s)$ and $\delta(t)$, then
$\frac{x^*}{2}$ is a valid correction vector for all $S\in\Sscr$, 
leading to $\bomc(p)\le\frac{3}{2}c(x^*)$,
which is \citegenitivs{Wol80} analysis of Christofides' algorithm for the classical metric TSP.
But in the $s$-$t$-path TSP we will in general have narrow cuts.

The following property is useful:

\begin{lemma}[\cite{AnKS12}]
\label{chain}
The narrow cuts form a chain: there are 
sets $\{s\}=L_0\subset L_1\subset \cdots \subset L_{\ell-1}\subset L_\ell=V\setminus\{t\}$
so that $\Cscr=\{\delta(L_i):i=0,\ldots,\ell\}$.
\end{lemma}

\prove
We have $x^*(\delta(s))=x^*(\delta(t))=1$ and hence $\delta(s)\in\Cscr$ and $\delta(V\setminus\{t\})=\delta(t)\in\Cscr$.
Suppose we have $U'\subset V$ and $U''\subset V$ with $\delta(U'),\delta(U'')\in\Cscr$ and
$s\in U'\cap U''\notin \{U',U''\}$. Then $t\notin U'\cup U''$ and
$2+2>x^*(\delta(U'))+x^*(\delta(U'')) \ge x^*(\delta(U'\setminus U'')) + x^*(\delta(U''\setminus U'))\ge 2+2$,
a contradiction.
\endproof

We remark:

\begin{proposition}
\label{computenarrowcuts}
Given $x^*$, the set $\Cscr$ of narrow cuts can be computed in polynomial time.
\end{proposition}

\prove
Lemma \ref{chain} implies that there is an order $V=\{v_1,v_2,\ldots,v_n\}$ such that each narrow cut
has the form $\delta(\{v_1,\ldots,v_i\})$ and is the only $\{v_i,v_{i+1}\}$-cut $C$ with $x^*(C)<2$.
Thus it suffices to compute a $\{v,w\}$-cut $C$ with minimum $x^*(C)$ for all $v,w\in V$ with $v\not=w$. 
This can be done by ${n\choose 2}$ applications of any polynomial-time max-flow algorithm 
(or more efficiently by computing a Gomory-Hu tree).
\endproof

Similarly to Lemma \ref{chain} we have:
\begin{lemma}
\label{smallintersection}
For all $C,C'\in\Cscr$ with $C\not=C'$ we have
$$x^*(C\cap C') \ \le \ \frac{1}{2} x^*(C) + \frac{1}{2} x^*(C') -1.$$
\end{lemma}

\prove
Let $C=\delta(L_i)$ and $C'=\delta(L_j)$ with $0\le i<j\le \ell$.
Then 
$x^*(C) + x^*(C') - 2 x^*(C\cap C') = x^*(\delta(L_j\setminus L_i)) \ge 2$.
\endproof

\newcommand{\peven}{p^C_{\text{\rm even}}}
\newcommand{\pone}{p^C_{\text{\rm one}}}
\newcommand{\pmany}{p^C_{\text{\rm many}}}

Moreover, \cite{AnKS12} and \cite{Seb13} observed that
\begin{equation}
\label{boundeventreesincut}
\peven \ := \ \Ssumeven p_S \ \le \ x^*(C)-1
\end{equation}
\vspace*{-2mm}
and
\begin{equation}
\label{boundonetreesincut}
\pone \ := \ \Ssumone p_S \ \ge \ 2-x^*(C)
\end{equation}
for every narrow cut $C\in\Cscr$.
Note that  (\ref{boundeventreesincut}) and (\ref{boundonetreesincut}) follow directly from
$x^*(C) = \sum_{S\in\Sscr}p_S |S\cap C|$ for all $C\subseteq E$ and $|S\cap C|\ge 1$ for all $C\in\Cscr$ and $S\in\Sscr$.
From this we also get
\begin{equation}
\label{boundmanytreesincut}
\pmany \ := \ \sum_{S\in\Sscr}p_S \left\lfloor \frac{|S\cap C|-1}{2} \right\rfloor 
\ = \ \frac{1}{2} \left( x^*(C)-1- \peven \right)
\end{equation}
for all $C\in\Cscr$.

\begin{figure}
\begin{center}
\begin{tikzpicture}[thick, minimum size=7, inner sep=2, scale=0.9]
  \node at (-6.4,1.2) {$s$};
  \node[circle,draw] (s) at (-6,1.2) {};
  \node[circle,draw, fill=grey] (a) at (-4,1.2) {};
  \node[circle,draw, fill=grey] (b2) at (-2,0) {};
  \node[circle,draw, fill=grey] (c) at (0,1.2) {};
  \node[circle,draw, fill=grey] (d2) at (2,2.4) {};
  \node[circle,draw] (d) at (2,1.2) {};
  \node[circle,draw, fill=grey] (e3) at (4,0) {};
  \node[circle,draw, fill=grey] (e) at (4,1.2) {};
  \node[circle,draw, fill=grey] (e2) at (4,2.4) {};
  \node[circle,draw] (f) at (6,1.2) {};
  \node[circle,draw] (f2) at (6,0) {};
  \node[circle,draw, fill=grey] (g) at (8,1.2) {};
  \node[circle,draw] (g2) at (8,2.4) {};
  \node[circle,draw] (h) at (10,1.2) {};
  \node at (10.4,1.2) {$t$};
\draw[red] (e) -- (e3);
\draw[turq] (e) -- (e2);
\draw[red] (d2) -- (d);
\draw[turq] (f) -- (f2);
\draw[red] (g) -- (g2);
\draw[turq] (s) -- (a);
\draw[red] (c) -- (d);
\draw[turq] (d2) -- (e2);
\draw[turq] (e) -- (f);
\draw[bend right,red] (a) to (b2);
\draw[out=40,in=180,turq] (a) to (d2);
\draw[red] (e2) to (g2);
\draw[out=0,in=220,turq] (f2) to (h);
\draw[grey, densely dotted] (-5,-0.2)--(-5,2.6);
\draw[grey] (-3,-0.2)--(-3,2.6);
\draw[grey, densely dotted] (-1,-0.2)--(-1,2.6);
\draw[grey] (1,-0.2)--(1,2.6);
\draw[grey, densely dotted] (3,-0.2)--(3,2.6);
\draw[grey] (5,-0.2)--(5,2.6);
\draw[grey] (7,-0.2)--(7,2.6);
\draw[grey, densely dotted] (9,-0.2)--(9,2.6);
\end{tikzpicture}
\end{center}
\caption{A spanning tree $(V,S)$. Filled circles show vertices in $T_S$.
The edge set $S$ is partitioned into its $s$-$t$-path $I_S$ (turquoise) and its $T_S$-join $J_S$ (red).
The narrow cuts are those shown in grey (dotted and solid).
Those that have an even number of edges of $S$ are shown in solid grey; 
each of them contains (at least) one red and (at least) one blue edge.
\label{introexample}}
\end{figure}
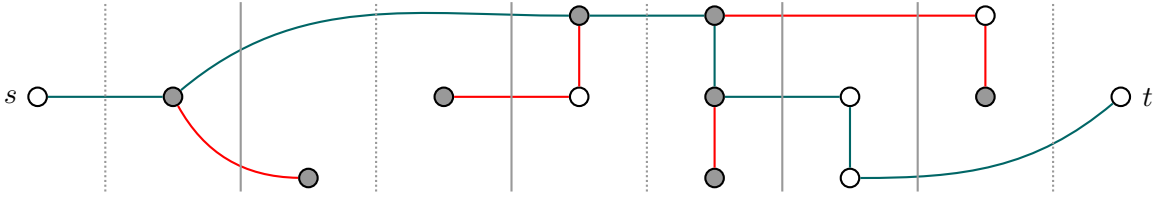

For $S\in\Sscr$ let $I_S$ denote the edge set of the $s$-$t$-path in $(V,S)$. 
Let $J_S:=S\setminus I_S$ and note that $J_S$ is a $T_S$-join.
See Figure \ref{introexample} for an example.
\cite{Seb13} observed that
\begin{equation}
\label{sebopacking}
\sum_{C\in\Cscr} \, \Ssumone p_S \chi^{S\cap C} 
\ \le \ \sum_{S\in\Sscr}p_S \chi^{I_S}
\end{equation}
because $I_S$ (in fact every $s$-$t$-path) intersects every narrow cut (in fact every $\{s,t\}$-cut),
and if $S\cap \delta(L_i) = S\cap\delta(L_j)=\{e\}$ for some $0\le i<j\le \ell$ and $e\in E$,
then $S\cap \delta(L_j\setminus L_i)=\emptyset$, contradicting that $(V,S)$ is connected.

\subsection{Correction vectors}

Consider
\begin{equation}
\label{correctionvector}
y^S \ := \ \beta x^* + (1-2\beta)\chi^{J_S} +  z^S
\end{equation}
for $S\in\Sscr$,
where $0\le\beta\le\frac{1}{2}$, and
$z^S\in\mathbb{R}^E_{\ge 0}$ is a nonnegative vector satisfying
\begin{equation}
\label{aksreq}
z^S(C) \ \ge \ \beta(2-x^*(C))
\end{equation}
for all 
$C\in\Cscr$ with $|S\cap C|$ even.

\begin{lemma}[\cite{AnKS12}]
For every $S\in\Sscr$ 
and every $T_S$-cut $C$ we have
$y^S(C)\ge 1$.
\end{lemma}

\prove 
Let $S\in\Sscr$ and $C$ be a $T_S$-cut. 
$J_S$ is a $T_S$-join, so $|J_S\cap C|\ge 1$.

If $C\notin\Cscr$, then $x^*(C)\ge 2$ and hence
$y^S(C)\ge \beta x^*(C) + (1-2\beta) |J_S\cap C| 
\ge 2\beta + 1-2\beta = 1$.

If $C\in\Cscr$, then 
$|I_S \cap C|$ (the intersection of an $s$-$t$-path and an $\{s,t\}$-cut) is odd and
$|J_S\cap C|$ (the intersection of a $T_S$-join and a $T_S$-cut) is odd,
so $|S\cap C|$ is even.
Hence
$y^S(C) =  \beta x^*(C) + (1-2\beta) |J_S\cap C| + z^S(C)
\ge \beta x^*(C) + 1-2\beta + z^S(C)\ge 1$ due to (\ref{aksreq}).
\endproof

So $y^S$ is in the $T_S$-join polyhedron (\ref{TSjoinpolyhedron}) for all $S\in\Sscr$.
For any distribution $p$ and nonnegative vectors $(z^S)_{S\in\Sscr}$ with (\ref{aksreq}),
we get with (\ref{basicbound}) and (\ref{correctionvector}) the bound
\begin{equation}
\label{BOMCcost}
\bomc(p) \ \le \ (1+\beta) c(x^*) + (1-2\beta) \sum_{S\in\Sscr} p_S c(J_S) 
+ \sum_{S\in\Sscr} p_S c(z^S).
\end{equation}
Now the question is how to choose the vectors $z^S$.

\subsection{The analyses of An, Kleinberg, Shmoys, and Seb\H{o} \label{oldanalysis}}

\cite{AnKS12}, \cite{Seb13}, and then also \cite{Gao15}, chose 
\begin{equation}
\label{oldchoiceofrS}
z^S \ = \ \alpha \chi^{I_S} 
+ \Csumeven  \max\left\{0,\, 2\beta - \alpha - \beta x^*(C) \big. \right\} v^C,
\end{equation}
where $v^C\in\mathbb{R}_{\ge 0}^E$ are vectors with $v^C(C)\ge 1$ for all $C\in\Cscr$.

As $I_S\cap C\not=\emptyset$ (recall that $I_S$ is the edge set of an $s$-$t$-path),
this choice implies $z^S(C)\ge \alpha+(2\beta-\alpha-\beta x^*(C)) = 2\beta -\beta x^*(C)$
for all $C\in\Cscr$ with $|S\cap C|$ even, as required in (\ref{aksreq}).

Writing $I_p:=\sum_{S\in\Sscr}p_S \chi^{I_S}$ and $J_p:=\sum_{S\in\Sscr}p_S \chi^{J_S}$
we get, with (\ref{BOMCcost}) and (\ref{oldchoiceofrS}) and $c(I_p)+c(J_p)=c(x^*)$:
\begin{eqnarray*}
\bomc(p) 
&\le& (1+\beta) c(x^*) + (1-2\beta) c(J_p) + \sum_{S\in\Sscr}p_S c(z^S) \\
&=&  (1+\beta) c(x^*) + (1-2\beta) c(J_p) 
+ \alpha c(I_p)  \\
&& \qquad\qquad
+ \sum_{S\in\Sscr} p_S \! \Csumeven  
\!\!\!\!\!\!\
\max\left\{0,\, 2\beta - \alpha - \beta x^*(C) \big.\right\} c(v^C) \\
&=& (1+\alpha+\beta) c(x^*) + (1-\alpha-2\beta) c(J_p) \\
&& \qquad\qquad
+ \sum_{C\in\Cscr} \,
\peven \max\left\{0,\, 2\beta - \alpha - \beta x^*(C)\right\} c(v^C) \\
&\le& (1+\alpha+\beta) c(x^*) + (1-\alpha-2\beta) c(J_p) \\
&& \qquad\qquad
+ \sum_{C\in\Cscr} (x^*(C)-1) \max\left\{ 0,\, 2\beta - \alpha - \beta x^*(C)\right\} c(v^C),
\end{eqnarray*}
where we used 
(\ref{boundeventreesincut}) in the last inequality.
The three papers choose the vectors $v^C$ ($C\in\Cscr$) differently.

\smallskip
\cite{AnKS12} showed that $v^C$ ($C\in\Cscr$) can be chosen so that $\sum_{C\in\Cscr}v^C \le x^*$.
Observing $(x-1) \left(2\beta - \alpha - \beta x \right) \le \frac{(\beta-\alpha)^2}{4\beta}$ for all $x\in\mathbb{R}$ and
setting $\beta=\frac{1}{\sqrt{5}}$ and $\alpha=1-\frac{2}{\sqrt{5}}$, 
they obtained $\bomc(p)\le \frac{1+\sqrt{5}}{2} c(x^*)$.
\smallskip

\cite{Seb13} chose
$v^C:=\frac{1}{2-x^*(C)}\Ssumone p_S \chi^{S\cap C}$ for $C\in\Cscr$.
Then indeed $v^C(C)\ge 1$ for all $C\in\Cscr$ due to  (\ref{boundonetreesincut}).
Using (\ref{sebopacking}) and
observing $\frac{(x-1) (2\beta - \alpha - \beta x)}{2-x}\le \alpha+\beta -2\sqrt{\alpha\beta}$ for all $x\le 2$ one gets
$\bomc(p) \le (1+\alpha+\beta) c(x^*) + (1-\alpha-2\beta) c(J_p) + 
(\alpha+\beta -2\sqrt{\alpha\beta}) c(I_p)$.
Setting  
$\beta=\frac{2}{5}$ and $\alpha=\frac{1}{10}$, this yields $\bomc(p)\le\frac{8}{5}c(x^*)$.
(\cite{Seb13} set $\beta=\frac{4}{9}$ and $\alpha=\frac{1}{9}$, which yields the same bound if 
$c(J_p)\ge\frac{3}{5}c(x^*)$; otherwise he used $y^S=\chi^{J_S}$ ($S\in\Sscr$) as correction vectors).

\cite{Gao15} simply chose $v^C$ to be the incidence vector of a cheapest edge in $C$,
which is clearly best possible in this framework, 
but he could not obtain a better approximation ratio.

\subsection{New approach}

We use the ideas of \cite{AnKS12} and \cite{Seb13} but define $z^S$ differently; see Section \ref{sectionnewvectors}.
Like \cite{Seb13}, we will bound $c(z^S)$ by a constant fraction of $c(I_p)$.
More precisely, we will find vectors $z^S\in\mathbb{R}_{\ge 0}^E$ with (\ref{aksreq})
and $\sum_{S\in\Sscr}p_S c(z^S) \le (1-2\beta) c(I_p)$. 
With (\ref{BOMCcost}) this will immediately yield 
$\bomc(p) \le (2-\beta) c(x^*)$.
The question is how large we can choose $\beta$.

In the Appendix we give an example that shows that it will in general not
be possible to choose $\beta>\frac{2}{5}$ and hence directly improve on \citegenitivs{Seb13} 
approximation ratio of $\frac{8}{5}$.
Therefore we will modify $p$ first.
By reassembling the trees that contribute to the convex combination $x^*=\sum_{S\in\Sscr}p_S \chi^S$
we eliminate the most critical configurations.
We will show how in Section \ref{sectionreassemble}.
Then we can complete the calculation in Section \ref{sectionfirstapx} and obtain an
improved approximation ratio.

\section{New correction vectors}
\label{sectionnewvectors}

Since we want to bound the weighted sum of the costs of the vectors $z^S$ by a multiple of $c(I_p)$
(the weighted sum of the costs of the $s$-$t$-paths $I_S$, $S\in\Sscr$), each pair $(S,e)$ with $S\in\Sscr$ and $e\in I_S$ 
will make a contribution to some of the vectors $z^{S'}$ (where $S'\in\Sscr$ can be $S$ or a different tree).
As in \citegenitivs{Seb13} analysis, there are two types of contributions.
A pair $(S,e)$ will contribute $\gamma_{S,e}(1-2\beta)$ to $z^S$
and a total of $(1-\gamma_{S,e})(1-2\beta)$ to vectors $z^{S'}$ for other trees $S'$.
The latter contribution is distributed as follows: if $C\in\Cscr$ is the narrow cut with
$S\cap C=\{e\}$ (there can be only one such cut), then $e$ will contribute to $z^{S'}$ for all
$S'\in\Sscr$ with $|S'\cap C|$ even.
\citegenitivs{Seb13} analysis is essentially equivalent to choosing $\gamma_{S,e}=\frac{1}{2}$ 
for all $S\in\Sscr$ and $e\in I_S$.
We will obtain an improvement by choosing individual values.

\subsection{The new vectors}

For any $S\in\Sscr$ and $C\in\Cscr$ choose an edge $e^S_C\in I_S\cap C$.
Moreover, for any $C\in\Cscr$ let $e_C\in C$ be a minimum cost edge in $C$.
For any pair $S\in\Sscr$ and $e\in I_S$ we will choose a number $0\le \gamma_{S,e}\le 1$ later.
Then, 
for $S\in\Sscr$, we set 
\begin{equation}
\label{defrS}
z^S \ := \ \sum_{e\in I_S} (1-2\beta) \gamma_{S,e} \chi^{\{e\}}
+ \Csumeven \!\!\!\!\!
\max\left\{0, \left( \beta(2-x^*(C)) - (1-2\beta)\gamma_{S,e^S_C} \right)\right\} \chi^{\{e_C\}}.
\end{equation}

The first term in (\ref{defrS})  is the direct contribution of the edges of $I_S$ to $z^S$.
The second term is exactly what is still needed to obtain (\ref{aksreq})
for all $C\in\Cscr$ with $|S\cap C|$ even; see the proof of Lemma \ref{boundfornewvectors} below.
We will have to show that the total cost of the weighted sum of the second terms in (\ref{defrS}) is not more than 
$\sum_{S\in\Sscr}p_S\sum_{e\in I_S} (1-2\beta) (1-\gamma_{S,e}) c(e)$; see Section \ref{sectionboundingcost}.

\begin{lemma}
\label{boundfornewvectors}
If $0\le\beta\le\frac{1}{2}$ and $\sum_{S\in\Sscr}p_S c(z^S) \le (1-2\beta) \sum_{S\in\Sscr}p_S c(I_S)$
for the vectors $z^S$ ($S\in\Sscr$) defined in {\rm(\ref{defrS})}, then
$\bomc(p) \le (2-\beta)c(x^*)$.
\end{lemma}

\prove
Note that the vectors $z^S$ are nonnegative for all $S\in\Sscr$ (as $\beta\le\frac{1}{2}$ and $\gamma_{S,e}\ge 0$ for all $e\in I_S$).
We have for every $S\in\Sscr$ and $C\in\Cscr$ with $|S\cap C|$ even: 
\vspace*{-2mm}
$$z^S(C) \ \ge \ (1-2\beta) \gamma_{S,e^S_C} + \left( \beta(2-x^*(C)) - (1-2\beta)\gamma_{S,e^S_C} \right) 
\ = \ \beta(2-x^*(C)),$$
as required by (\ref{aksreq}).
So the bound immmediately follows from (\ref{BOMCcost}), using
$\sum_{S\in\Sscr}p_S c(z^S) \le (1-2\beta) \sum_{S\in\Sscr}p_S c(I_S)$ and $\sum_{S\in\Sscr}p_S(c(J_S)+c(I_S))=c(x^*)$.
\endproof

We will try to maximize $\beta$.

\subsection{Bounding the cost \label{sectionboundingcost}}

The cost of the vectors $z^S$ ($S\in\Sscr$)
will of course depend on the choice of the $\gamma_{S,e}$ ($S\in\Sscr$, $e\in I_S$).
The desired inequality $\sum_{S\in\Sscr} p_S c(z^S) \le (1-2\beta) \sum_{S\in\Sscr}p_S c(I_S)$ is implied by
$$
 \Ssumeven \! p_S \max \left\{ 0, \, \beta(2-x^*(C)) - (1-2\beta) \gamma_{S, e^S_C} \right\} 
\ \le \ \!  \Ssumone \! p_S (1-2\beta)(1-\gamma_{S, e^S_C}),
$$
as may be seen from (\ref{defrS}) and will be formally shown in the proof of Lemma \ref{lemmaneedbenefit} below. 
To write this inequality in a more compact form (see (\ref{enoughbenefit}) below), we divide by $1-2\beta$ and use the following notation:

\begin{definition}
\label{defbenefit}
Given numbers $\gamma_{S,e}\ge 0$ for $S\in\Sscr$ and $e\in I_S$,
we define the \emph{benefit} of $(S,C)\in\Sscr\times\Cscr$ to be
$b_{S,C}:=\min\left\{\frac{\beta (2-x^*(C)) }{1-2\beta}, \gamma_{S,e^S_C} \right\}$
if $|S\cap C|$ is even,
$b_{S,C}:=1-\gamma_{S,e^S_C}$ 
if $|S\cap C|=1$,
and $b_{S,C}=0$ otherwise.
\end{definition}

Now all we need to show is that we have enough benefit at every narrow cut:

\begin{lemma}
\label{lemmaneedbenefit}
Let $0\le\beta<\frac{1}{2}$ and $0\le \gamma_{S,e}\le 1$ for $S\in\Sscr$ and $e\in I_S$. 
If
\begin{equation}
\label{enoughbenefit}
\sum_{S\in\Sscr} p_S b_{S,C} \ \ge \ \frac{\beta (2-x^*(C)) \peven}{1-2\beta}
\end{equation}
for all $C\in\Cscr$,
then $\bomc(p) \le \ (2-\beta) c(x^*)$.
\end{lemma}

\prove
By (\ref{enoughbenefit}) and Definition \ref{defbenefit} we have
 for every $C\in\Cscr$:
\begin{equation}
\label{enoughbenefitreformulated}
\frac{\beta (2-x^*(C)) \peven}{1-2\beta} \ \le 
 \Ssumeven
 \!\!\!\!\!\!\!\!\!\!\!
 p_S \min\left\{\frac{\beta (2-x^*(C))}{1-2\beta}, \gamma_{S,e^S_C} \right\}
+ \! \Ssumone
 \!\!\!\!\!\!\!\!\!
p_S (1-\gamma_{S,e^S_C}).
\end{equation}

We compute:
\begin{eqnarray*}
\frac{\sum_{S\in\Sscr} p_S c(z^S)}{1-2\beta}\!
&=&
\sum_{S\in\Sscr} p_S \left(
\sum_{e\in I_S} \gamma_{S,e} c(e)
+ \!\! \Csumeven
 \!\!\!\!\!\!\!\!\!
\max\left\{0, \left( \frac{\beta(2-x^*(C))}{1-2\beta} - \gamma_{S,e^S_C} \right)\right\} c(e_C)
\! \right) \\
&=&
\sum_{S\in\Sscr} p_S 
\sum_{e\in I_S} \gamma_{S,e} c(e)
\\
&&
+
\sum_{C\in\Cscr} \left( \!
\frac{\beta (2-x^*(C)) \peven}{1-2\beta}  
- \!\! \Ssumeven
 \!\!\!\!\!\!\!\!\!
p_S \min\left\{ \frac{\beta(2-x^*(C))}{1-2\beta}, \gamma_{S,e^S_C} \right\} 
\right)  c(e_C) 
\\
&\le&
\sum_{S\in\Sscr} p_S 
\sum_{e\in I_S} \gamma_{S,e} c(e)
+ 
\sum_{C\in\Cscr} \,
\Ssumone
 \!\!\!\!\!\!\!\!\!
p_S (1-\gamma_{S,e^S_C}) c(e_C)
\\
&\le&
\sum_{S\in\Sscr} p_S 
\sum_{e\in I_S} \gamma_{S,e} c(e)
+ 
\sum_{C\in\Cscr} \,
\Ssumone
p_S (1-\gamma_{S,e^S_C}) c(e^S_C)
\\
&\le&
\sum_{S\in\Sscr} p_S 
\sum_{e\in I_S} \gamma_{S,e} c(e)
+ 
\sum_{S\in\Sscr} p_S
\sum_{e\in I_S} 
(1-\gamma_{S,e}) c(e)
\\
&=&
\sum_{S\in\Sscr} p_S c(I_S)
\end{eqnarray*}
(we used 
(\ref{enoughbenefitreformulated}) in the first,
$\gamma_{S,e^S_C}\le 1$ and $c(e_C)\le c(e^S_C)$ in the second,
and (\ref{sebopacking}) in the third inequality).
Now the assertion follows from Lemma \ref{boundfornewvectors}.
\endproof

Let us quickly check the obvious (although we will not need it):
if we simply choose $\gamma_{S,e}=\frac{1}{2}$ for all $S\in\Sscr$ and $e\in I_S$,
then we have for all $C\in\Cscr$:
if $ \frac{\beta}{1-2\beta} (2-x^*(C)) \le \frac{1}{2}$, then
$\sum_{S\in\Sscr} p_S b_{S,C} \ge \frac{\beta}{1-2\beta} (2-x^*(C)) \peven$, and otherwise
$$\sum_{S\in\Sscr} p_S b_{S,C}
 \ = \ \pone  \frac{1}{2} + \peven \frac{1}{2}
 \ \ge \ \frac{1}{2}((2-x^*(C)) +\peven)
 \ \ge \ 2 (2-x^*(C)) \peven 
$$
(the first inequality follows from (\ref{boundonetreesincut}), and the second one 
follows from $0\le (2-x^*(C))+\peven\le 1$ (cf.\ (\ref{boundeventreesincut})) and
the fact that $a+b\ge 4ab$ for all $a,b\ge 0$ with $a+b\le 1$).

So we have (\ref{enoughbenefit}) for $\beta=\frac{2}{5}$ and obtain again
\citegenitivs{Seb13} approximation ratio $\frac{8}{5}$. 
This is tight for $x^*(C)=\frac{3}{2}$ and $\pone=\peven=\frac{1}{2}$, when the total benefit $\sum_{S\in\Sscr} p_S b_{S,C}$ is $\frac{1}{2}$.

\subsection{More benefit \label{morebenefit}}

Let us first give an informal description of our idea. 
The approximation guarantee can be readily improved if there are no cuts $C$ with $x^*(C)\approx\frac{3}{2}$
(and $\pone\approx\frac{1}{2}$ and $\peven\approx\frac{1}{2}$; see the previous paragraph).
We would like to obtain more benefit (i.e., more than $\frac{1}{2}$) for ``critical'' cuts $C$ (i.e., those with $x^*(C)\approx\frac{3}{2}$),
and to achieve this we will reduce the benefit for less critical cuts (but still have enough).
If $C$ is a critical cut, $S\in\Sscr$, and $e^S_C$ does not belong to any other critical cut,
we can choose $\gamma_{S,e}>\frac{1}{2}$ if $|S\cap C|$ is even and $\gamma_{S,e}<\frac{1}{2}$ if $|S\cap C|=1$.
We will describe how exactly at the beginning of Section \ref{sectionfirstapx}.
In this case we get $b_{S,C}>\frac{1}{2}$.

However, this does not work if, for a critical cut $C$ and every $S\in\Sscr$, 
the edge $e^S_C$ also belongs to another critical cut $C'$ (and $|S\cap C|+|S\cap C'|=3$).
In this case we can only choose $\gamma_{S,e}=\frac{1}{2}$ and get $b_{S,C}=\frac{1}{2}$.
But note that then $S$ has at least one edge in $C\cap C'$.
However, due to Lemma \ref{smallintersection}, there cannot be too many edges 
in the intersection of two critical cuts: certainly less than one per tree on average.

We will fix a constant $\xi$ between $\frac{3}{2}$ and 2, and consider 
$\Cscr^{\xi}:=\{C\in\Cscr: x^*(C)<\xi\}$. 
These cuts were called $(\xi-1)$-narrow by \cite{AnKS12}, but we prefer to call them \emph{$\xi$-narrow}.
Cuts that are not $\xi$-narrow will not be critical.
For any $C\in\Cscr^{\xi}\setminus\{\delta(s),\delta(t)\}$, let 
$C_{\leftarrow}$ and $C_{\rightarrow}$ be the adjacent $\xi$-narrow cuts in both directions. 
More precisely, if $C=\delta(L_j)$, then 
$C_{\leftarrow}:=\delta(L_i)$ and $C_{\rightarrow}:=\delta(L_k)$, where 
$i=\max\{\iota: \iota<j,\, x^*(\delta(L_{\iota}))<\xi\}$ and
$k=\min\{\kappa: \kappa>j,\, x^*(\delta(L_{\kappa}))<\xi\}$ (cf.\ Lemma \ref{chain}).

Let $C$ be a critical cut.
By Lemma \ref{smallintersection}, $x^*(C\cap C_{\leftarrow})<1$ and $x^*(C\cap C_{\rightarrow})<1$,
so there must be trees $S\in\Sscr$ with $p_S>0$ with less than two edges in the disjoint union $(C\cap C_{\leftarrow}) \cupp (C\cap C_{\rightarrow})$.
Therefore, if every tree $S$ with $p_S>0$ has either larger benefit
than $\frac{1}{2}$ or
(benefit exactly $\frac{1}{2}$ and at least
two edges in $(C\cap C_{\leftarrow}) \cupp (C\cap C_{\rightarrow})$), 
we get $\sum_{S\in\Sscr} p_S b_{S,C} > \frac{1}{2}$, which leads to an improvement. 
This will be essentially our argument.

\newcommand{\good}{\textsc{good}}

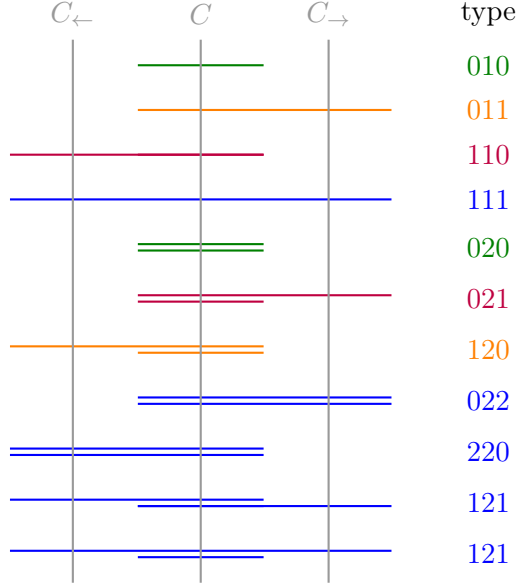
\begin{figure}[bht]
\begin{center}
\begin{tikzpicture}[thick, minimum size=0, inner sep=0, scale=0.85]

  \node[circle] (c2) at (-3,6.3) {};
  \node[circle] (c3) at (-3,5.6) {};
  \node[circle] (c6) at (-3,3.3) {};
  \node[circle] (c8) at (-3,1.7) {};
  \node[circle] (c81) at (-3,1.6) {};
  \node[circle] (c9) at (-3,0.9) {};
  \node[circle] (c10) at (-3,0.1) {};
  \node[circle] (d0) at (-1,7.7) {};
  \node[circle] (d1) at (-1,7.0) {};
  \node[circle] (d2) at (-1,6.3) {};
  \node[circle] (d4) at (-1,4.9) {};
  \node[circle] (d41) at (-1,4.8) {};
  \node[circle] (d5) at (-1,4.1) {};
  \node[circle] (d51) at (-1,4.0) {};
  \node[circle] (d61) at (-1,3.2) {};
  \node[circle] (d7) at (-1,2.5) {};
  \node[circle] (d71) at (-1,2.4) {};
  \node[circle] (d91) at (-1,0.8) {};
  \node[circle] (d101) at (-1,0.0) {};
  \node[circle] (e0) at (1,7.7) {};
  \node[circle] (e2) at (1,6.3) {};
  \node[circle] (e4) at (1,4.9) {};
  \node[circle] (e41) at (1,4.8) {};
  \node[circle] (e51) at (1,4.0) {};
  \node[circle] (e6) at (1,3.3) {};
  \node[circle] (e61) at (1,3.2) {};
  \node[circle] (e8) at (1,1.7) {};
  \node[circle] (e81) at (1,1.6) {};
 \node[circle] (e9) at (1,0.9) {};
  \node[circle] (e91) at (1,0.8) {};
  \node[circle] (e101) at (1,0.0) {};
   \node[circle] (f1) at (3,7.0) {};
 \node[circle] (f3) at (3,5.6) {};
 \node[circle] (f5) at (3,4.1) {};
 \node[circle] (f7) at (3,2.5) {};
  \node[circle] (f71) at (3,2.4) {};
 \node[circle] (f91) at (3,0.8) {};
  \node[circle] (f10) at (3,0.1) {};
 \node[darkgreen] (g0) at (4.5,7.7) {010};
 \node[orange] (g1) at (4.5,7.0) {011};
 \node[purple] (g2) at (4.5,6.3) {110};
 \node[blue] (g3) at (4.5,5.6) {111};
 \node[darkgreen] (g4) at (4.5,4.85) {020};
 \node[purple] (g5) at (4.5,4.05) {021};
 \node[orange] (g6) at (4.5,3.25) {120};
 \node[blue] (g7) at (4.5,2.45) {022};
 \node[blue] (g8) at (4.5,1.65) {220};
 \node[blue] (g9) at (4.5,0.85) {121};
 \node[blue] (g10) at (4.5,0.05) {121};

\draw[darkgreen] (d0) -- (e0);
\draw[orange] (d1) -- (f1);
\draw[purple] (c2) -- (e2);
\draw[purple] (d2) -- (e2);
\draw[blue] (c3) -- (f3);
\draw[darkgreen] (d4) -- (e4);
\draw[darkgreen] (d41) -- (e41);
\draw[purple] (d5) -- (f5);
\draw[purple] (d51) -- (e51);
\draw[orange] (c6) -- (e6);
\draw[orange] (d61) -- (e61);
\draw[blue] (d7) -- (f7);
\draw[blue] (d71) -- (f71);
\draw[blue] (c8) -- (e8);
\draw[blue] (c81) -- (e81);
\draw[blue] (c9) -- (e9);
\draw[blue] (d91) -- (f91);
\draw[blue] (d91) -- (e91);
\draw[blue] (c10) -- (f10);
\draw[blue] (d101) -- (e101);
\draw[grey] (-2,-0.4)--(-2,8.1);
\draw[grey] (0,-0.4)--(0,8.1);
\draw[grey] (2,-0.4)--(2,8.1);
\node[grey] at (-2,8.5) {$C_{\leftarrow}$};
\node[grey] at (0,8.5) {$C$};
\node[grey] at (2,8.5) {$C_{\rightarrow}$};
\node at (4.5,8.5) {type};
\end{tikzpicture}
\end{center}
\caption{Configurations at a $\xi$-narrow cut $C$; here $C_{\leftarrow}$ and $C_{\rightarrow}$ are the adjacent $\xi$-narrow cuts.
The types, according to Definition \ref{types}, are shown on the right.
In each configuration we see the (one or two) edges of $S\cap C$ for some $S\in\Sscr$; 
if there are two, the edge $e^S_C$ that belongs to the $s$-$t$-path is shown on top.
If $C$ is critical (i.e., $x^*(C)\approx\frac{3}{2}$), each of these configurations will get benefit at least $\frac{1}{2}$.
Green configurations will get benefit strictly more than $\frac{1}{2}$, which is possible because $e^S_C$ belongs to no other critical cut.
Blue configurations have two edges in $(C\cap C_{\leftarrow})\cupp(C\cap C_{\rightarrow})$,
which is more than possible on average.
Orange and purple configurations are problematic.
Configurations with even more edges in $(C\cap C_{\leftarrow})\cupp(C\cap C_{\rightarrow})$ or
with three or more edges in $C$ (these will be called \good) are not shown.
\label{configurations}}
\end{figure}

As can be seen from Figure \ref{configurations}, there
are however ---in addition to trees with more than two edges in $C$--- 
four configurations (orange and purple) that do not have this property.
Our reassembling step, to be described next, aims at avoiding two out of these four configurations
(one of the two orange ones and one of the two purple ones). It will turn out that this is enough.

\section{Reassembling Trees \label{sectionreassemble}}

In this section we show how to reassemble the trees that contribute to the convex combination
$x^*=\sum_{S\in\Sscr} p_S \chi^S$ in order to remove certain bad configurations.

Given a constant $\frac{3}{2}<\xi<2$ (the exact value will be chosen later), 
we number the $\xi$-narrow cuts $\Cscr^{\xi}=\{C_0,\ldots,C_{\ell'}\}$ from left to right; i.e., 
for $0\le i<k\le\ell'$ and $C_{i}=\delta(L_{\iota})$ and $C_{k}=\delta(L_{\kappa})$ we have $\iota<\kappa$.
Note that $C_0=\delta(s)$ and $C_{\ell'}=\delta(t)$.
Moreover, for $C=C_j$ ($i\in\{1,\ldots,\ell'-1\}$) we have $C_{\leftarrow}=C_{j-1}$ and $C_{\rightarrow}=C_{j+1}$.

\subsection{Types of trees at \boldmath{$\xi$}-narrow cuts}

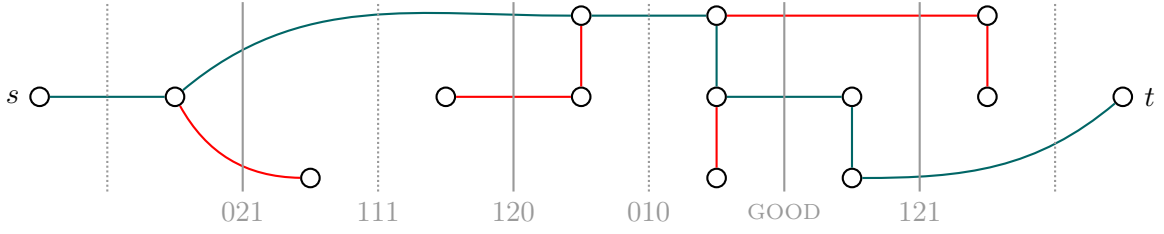
\begin{figure}[bht]
\begin{center}
\begin{tikzpicture}[thick, minimum size=7, inner sep=2, scale=0.9]
  \node at (-6.4,1.2) {$s$};
  \node[circle,draw] (s) at (-6,1.2) {};
  \node[circle,draw] (a) at (-4,1.2) {};
  \node[circle,draw] (b2) at (-2,0) {};
  \node[circle,draw] (c) at (0,1.2) {};
  \node[circle,draw] (d2) at (2,2.4) {};
  \node[circle,draw] (d) at (2,1.2) {};
  \node[circle,draw] (e3) at (4,0) {};
  \node[circle,draw] (e) at (4,1.2) {};
  \node[circle,draw] (e2) at (4,2.4) {};
  \node[circle,draw] (f) at (6,1.2) {};
  \node[circle,draw] (f2) at (6,0) {};
  \node[circle,draw] (g) at (8,1.2) {};
  \node[circle,draw] (g2) at (8,2.4) {};
  \node[circle,draw] (h) at (10,1.2) {};
  \node at (10.4,1.2) {$t$};
\draw[red] (e) -- (e3);
\draw[turq] (e) -- (e2);
\draw[red] (d2) -- (d);
\draw[turq] (f) -- (f2);
\draw[red] (g) -- (g2);
\draw[turq] (s) -- (a);
\draw[red] (c) -- (d);
\draw[turq] (d2) -- (e2);
\draw[turq] (e) -- (f);
\draw[bend right,red] (a) to (b2);
\draw[out=40,in=180,turq] (a) to (d2);
\draw[red] (e2) to (g2);
\draw[out=0,in=220,turq] (f2) to (h);
\draw[grey, densely dotted] (-5,-0.2)--(-5,2.6);
\draw[grey] (-3,-0.2)--(-3,2.6);
\draw[grey, densely dotted] (-1,-0.2)--(-1,2.6);
\draw[grey] (1,-0.2)--(1,2.6);
\draw[grey, densely dotted] (3,-0.2)--(3,2.6);
\draw[grey] (5,-0.2)--(5,2.6);
\draw[grey] (7,-0.2)--(7,2.6);
\draw[grey, densely dotted] (9,-0.2)--(9,2.6);
\node[grey] at (-3,-0.5) {021};
\node[grey] at (-1,-0.5) {111};
\node[grey] at (1,-0.5) {120};
\node[grey] at (3,-0.5) {010};
\node[grey] at (5,-0.5) {\good};
\node[grey] at (7,-0.5) {121};
\end{tikzpicture}
\end{center}
\caption{The spanning tree $(V,S)$ from Figure \ref{introexample}. Assuming that all narrow cuts (grey vertical lines) are $\xi$-narrow,
we list the type of $S$ at each $C\in\Cscr^{\xi}\setminus\{\delta(s),\delta(t)\}$ according to Definition \ref{types}.
For the cut $C$ marked {\good} we have $l=0$, $m=2$, $r=1$, and $S\cap C'\not=\{e\}$ for all $e\in S\cap C$ and all $C'\in\Cscr^{\xi}$.
\label{treeclasses}}
\end{figure}

We now define the type of a spanning tree at a $\xi$-narrow cut. We remark that
the type can depend on the value of $\xi$. This is no problem since $\xi$ is a fixed constant (chosen later).

\begin{definition}
\label{types}
Fix a constant $\frac{3}{2}<\xi<2$.
Then, for any tree $S\in\Sscr$ and any cut $C\in\Cscr^{\xi}\setminus\{\delta(s),\delta(t)\}$ 
we define the \emph{type} of $S$ at $C$ as follows.
Let $l=|S\cap C\cap C_{\leftarrow}|$ and $m=|S\cap C|$ and $r=|S\cap C\cap C_{\rightarrow}|$.
If $m\ge 3$ or $l+r\ge 3$ or $(l+r\ge 1$ and $S\cap C'\not=\{e\}$ for all $e\in S\cap C$ and all $C'\in\Cscr^{\xi})$,
then the type is ``\;\!\good'', otherwise the type is ``\;\!$lmr$''.
 \end{definition}
 
 Note that if the type of $S$ at $C$ is \good, then $|S\cap C|\ge 2$.
 See Figure \ref{configurations} for a list of all types that are not \good, and Figure \ref{treeclasses} for the types
 of the tree in Figure \ref{introexample} (assuming that all narrow cuts are in $\Cscr^{\xi}$). 
 All types are also listed in Table \ref{benefittable} on page \pageref{benefittable}.

\subsection{Reassembling lemma}

Here is our key lemma for reassembling trees.

\begin{lemma}
\label{reassemblelemma}
Let $i\in\{1,\ldots,\ell'-1\}$, so $C_i\in\Cscr^{\xi}\setminus\{\delta(s),\delta(t)\}$.
Let $S_1,S_2\in\Sscr$ 
such that $S_1$ has type 120 at $C_i$ and $S_2$ has type 011 at $C_i$.

Then there are two edges $e_1\in S_1$ 
and $e_2\in S_2$
such that
$S_1' := (S_1\setminus \{e_1\})\cup\{e_2\}\in\Sscr$ and
$S_2' := (S_2\setminus \{e_2\})\cup\{e_1\}\in\Sscr$.
Moreover we have:
\begin{itemize}
\item[\rm (a)] 
At each of $C_1,\ldots,C_{i-1}$, the new trees have the same type as the old trees.
\item[\rm (b)] 
At $C_i$, 
$S_1'$ has type 121, and
$S_2'$ has type 010.
\item[\rm (c)] 
If $S_1'$ has type 110 or 021 at $C_j$ for some $j>i$,
then $S_1$ has the same type at $C_j$.
\item[\rm (d)] 
If $S_2'$ has type 110 or 021 at $C_j$ for some $j>i$, then $S_2$ has the same type at $C_j$ or
$S_1'$ has type \good{} at each of $C_{i+1},\ldots,C_j$.
\end{itemize}
\end{lemma}

\begin{figure}
\begin{center}
\begin{tikzpicture}[thick, minimum size=6, inner sep = 0, scale=0.63]

\begin{scope};
  \node[circle,draw] (a1) at (-4,2) {};
  \node[circle,draw] (b1) at (-2,3) {};
  \node[circle,draw] (c1) at (0,3) {};
  \node[circle,draw] (d1) at (2,2) {};
  \node[circle,draw] (e1) at (4,2) {};
  \node[circle,draw] (f1) at (6,3) {};
\draw[red] (b1) -- (c1);
\draw[red] (a1) -- (d1);
\draw[red] (d1) -- (e1);
\draw[red] (e1) -- (f1);
\draw[red, densely dotted] (c1) -- (d1);
\node[red] at (-2.6,2.3) {\small $e_0$};
\node[red] at (1.35,2.65) {\small $e_1$};
  \node[circle,draw] (a2) at (-4,0) {};
  \node[circle,draw] (b2) at (-2,1) {};
  \node[circle,draw] (c2) at (0,1) {};
  \node[circle,draw] (d2) at (2,0) {};
  \node[circle,draw] (e2) at (4,0) {};
  \node[circle,draw] (f2) at (6,1) {};
\draw[turq] (a2) -- (c2);
\draw[turq] (b2) -- (a2);
\draw[turq] (d2) -- (e2);
\draw[turq] (d2) -- (f2);
\draw[turq, densely dotted] (c2) -- (f2);
\node[turq] at (1.35,1.3) {\small $e_2$};

\draw[grey, densely dotted] (-3,-0.3)--(-3,3.3);
\node[grey] at (-3,-0.7) {\small $C_h$};
\draw[grey, densely dotted] (-1,-0.3)--(-1,3.3);
\draw[grey] (1,-0.3)--(1,3.3);
\node[grey] at (1,-0.7) {\small $C_i$};
\draw[grey, densely dotted] (3,-0.3)--(3,3.3);
\draw[grey, densely dotted] (5,-0.3)--(5,3.3);
\node[grey] at (5,-0.7) {\small $C_k$};

\node[red] at (7,2.5) {\small $S_1$};
\node[turq] at (7,0.5) {\small $S_2$};
\end{scope}

\draw[->] (7.7,1.5) -- (8.7,1.5);

\begin{scope}[shift={(14.4,0)}];
\node[red] at (-5,2.5) {\small $S_1\!\!\textcolor{turq}{'}$};
\node[turq] at (-5,0.5) {\small $S_2\!\!\textcolor{red}{'}$};

  \node[circle,draw] (a1) at (-4,2) {};
  \node[circle,draw] (b1) at (-2,3) {};
  \node[circle,draw] (c1) at (0,3) {};
  \node[circle,draw] (d1) at (2,2) {};
  \node at (2.3,2.4) {\small $v_0$};
  \node[circle,draw] (e1) at (4,2) {};
  \node[circle,draw] (f1) at (6,3) {};
  \node at (6.3,3.4) {\small $v_2$};
\draw[red] (b1) -- (c1);
\draw[red] (a1) -- (d1);
\draw[red] (d1) -- (e1);
\draw[red] (e1) -- (f1);
  \node[circle,draw] (a2) at (-4,0) {};
  \node[circle,draw] (b2) at (-2,1) {};
  \node[circle,draw] (c2) at (0,1) {};
  \node[circle,draw] (d2) at (2,0) {};
  \node[circle,draw] (e2) at (4,0) {};
  \node[circle,draw] (f2) at (6,1) {};
\draw[turq] (a2) -- (c2);
\draw[turq] (b2) -- (a2);
\draw[turq] (d2) -- (e2);
\draw[turq] (d2) -- (f2);
\draw[turq, densely dotted] (c1) -- (f1);
\draw[red, densely dotted] (c2) -- (d2);
\node[red] at (-2.6,2.3) {\small $e_0$};
\node[red] at (1.35,0.65) {\small $e_1$};
\node[turq] at (1.35,3.3) {\small $e_2$};

\draw[grey, densely dotted] (-3,-0.3)--(-3,3.3);
\node[grey] at (-3,-0.7) {\small $C_h$};
\draw[grey, densely dotted] (-1,-0.3)--(-1,3.3);
\draw[grey] (1,-0.3)--(1,3.3);
\draw[grey, densely dotted] (3,-0.3)--(3,3.3);
\draw[grey, densely dotted] (5,-0.3)--(5,3.3);
\node[grey] at (1,-0.7) {\small $C_i$};
\node[grey] at (5,-0.7) {\small $C_k$};
\end{scope}

\end{tikzpicture}
\end{center}
\caption{Reassembling trees: exchanging two edges in trees of types \textcolor{red}{120} 
and \textcolor{turq}{011} (Lemma \ref{reassemblelemma}).
\label{patch1}}
\end{figure}
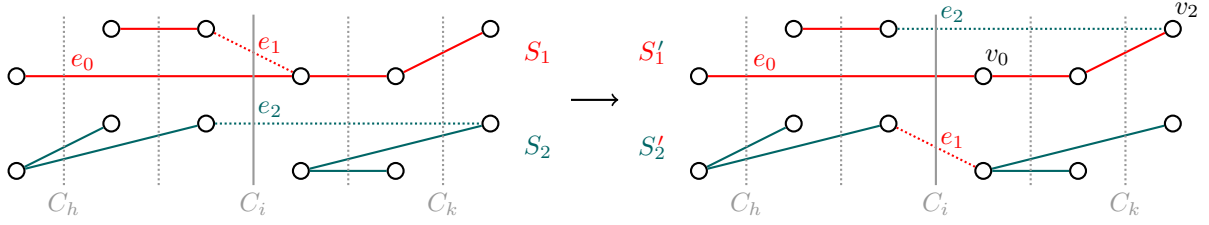

See Figure \ref{patch1} for an example.
\bigskip

\prove
%
Let $e_2$ be the only edge in $S_2\cap C_i$. 
As $S_2$ has type 011 at $C_i$, 
there is an index $k\in\{i+1,\ldots,\ell-1\}$ such that $e_2$
belongs to the cuts $C_i,\ldots,C_k$ but neither $C_{i-1}$ nor $C_{k+1}$.

Let $S_1\cap C_i=\{e_0,e_1\}$, where $e_0$ belongs to the $s$-$t$-path in $S_1$ (so $I_{S_1}\cap C_i = \{e_0\}$).
As $S_1$ has type 120 (and not \good) at $C_i$, there must be an edge in $S_1\cap C_i$ 
that is the only edge of $S_1$ in some other $\xi$-narrow cut, and
such an edge must belong to the $s$-$t$-path in $S_1$.
So there exists an index
$h\in\{0,\ldots,i-1\}$ such that $C_h\cap S_1=\{e_0\}$;
moreover $e_1\notin C_{i-1}\cup C_{i+1}$.

The graph $(V,S_1\cup\{e_2\})$ contains a circuit $A$.
Any circuit has even intersection with any cut.
As $e_2\notin C_h$, the circuit $A$ does not contain $e_0$.
However, $A$ must contain (at least) a second edge in $C_i$ besides $e_2$; so $A$ contains $e_1$. 

We have $S_1'\in\Sscr$ because $e_1$ belongs to the circuit $A$ in  $(V,S_1\cup\{e_2\})$.
We have $S_2'\in\Sscr$ because $S_2\cap C_i=\{e_2\}$ and $e_1\in C_i$. 

Property (a) is obvious as nothing changes for $C_0,\ldots,C_{i-1}$: note that $e_1,e_2\in C_i\setminus C_{i-1}$.
Property (b) follows from $S_1'\cap C_i=\{e_0,e_2\}$ and
$S_2'\cap C_i=\{e_1\}$.

To show (c) and (d), we first prove:

\noindent{\bf Claim:} $S_1'$ has type \good{} at $C_{i+1},\ldots,C_{k}$.

To this end, observe that $|S_1'\cap C_j|\ge |(S_1\setminus\{e_1\})\cap C_j|+1\ge 2$ for $j=i,\ldots,k$.

Let $v_2$ be the ``right'' endpoint of $e_2$, and $v_0$ the ``right'' endpoint of $e_0$
(i.e., if $C_i=\delta(L_{\iota})$, then $v_0,v_2\notin L_{\iota}$; cf.\ Figure \ref{patch1}.)
Let $P$ be the $v_0$-$v_2$-path in $S_1'$. 
We observe that (\stern) $P$ crosses 
every cut among $C_{i+1},\ldots,C_k$ an odd number of times and
every cut $C_j$ with $j\le i$ or $j>k$ an even number of times.
Therefore $P$ contains neither $e_0$ (it contains an even number of edges from $S_1'\cap C_h=\{e_0\}$)
nor $e_2$ (it contains an even number of edges from $S_1'\cap C_i=\{e_0,e_2\}$).

Now let $j\in\{i+1,\ldots,k\}$ and $e\in I_{S'_1}\cap C_j$. 
We have $e\not=e_0$ because $e_0\notin C_j$. 
Moreover, $e\not=e_2$ because $e_2\notin I_{S_1'}$ 
(as this $s$-$t$-path contains an odd number of edges from $S_1'\cap C_h=\{e_0\}$ and from
$S_1'\cap C_i=\{e_0,e_2\}$).
So $e\notin C_i$.

Suppose $S'_1\cap C_q=\{e\}$ for some $q$ (otherwise $S'_1$ has type \good{} at $C_j$ because $e_2$ guarantees $l+r\ge 1$); 
then $q>k$ (as $e_2\in C_i\cup\cdots \cup C_k$), and from (\stern) we get $e\notin E(P)$.
Then we have $e,e_2\in S'_1\cap C_j$ and $|E(P)\cap C_j|$ is odd and $e,e_2\notin E(P)$.
This implies $|S'_1\cap C_j|\ge 3$, so again $S'_1$ has type \good{} at $C_j$.
The claim is proved.

The claim directly implies (c) and (d) for $j\in\{i+1,\ldots,k\}$.

Now let $j\in\{k+1,\ldots,\ell'-1\}$.
Suppose $S_1$ and $S_1'$ have different types at $C_j$.
Since $l,m,r$ (cf.\ Definition \ref{types}) are identical at $C_j$, the only remaining reason for different types is 
that $S_1\cap C_{j'} = \{e\}$ (and thus $S_1'\cap C_{j'} = \{e,e_2\}$) 
for some $i<j'\le k$ and $e\in S_1\cap C_j=S_1'\cap C_j$. But then the new type is \good{}. 

Now suppose $S_2$ and $S_2'$ have different types at $C_j$ for some $j\in\{k+1,\ldots,\ell'-1\}$.
Again $l,m,r$ are identical at $C_j$, and the only remaining reason for different types is 
that $S_2'\cap C_{j'} = \{e\}$ (and thus $S_2\cap C_{j'} = \{e,e_2\}$) 
for some $i<j'\le k$ and $e\in S_2\cap C_j=S_2'\cap C_j$.
But then $e\notin I_{S_2}$, because $I_{S_2}$ contains an odd number of edges in $C_{j'}$ and must contain $e_2$.
This implies that $S_2\cap C_j$ contains another edge in addition to $e$, and hence $|S'_2\cap C_j|=|S_2\cap C_j|\ge 2$.
Therefore the new type can only be 120 or 121 or 220. 
\endproof

\subsection{Resulting types}

 \newcommand{\pgood}{p^C_{\text{\sc good}}}
 \newcommand{\p}[1]{p^C_{\text{\rm #1}}}
 \newcommand{\barpgood}{\bar p^{\,C}_{\text{\sc good}}}
 \newcommand{\barp}[1]{\bar p^{\,C}_{\text{\rm #1}}}
 \newcommand{\dbarpgood}{\bar{\bar p}^{\,C}_{\text{\sc good}}}
 \newcommand{\dbarp}[1]{\bar{\bar p}^{\,C}_{\text{\rm #1}}}
 
 Let $\p{$\tau$}:=\sum_{S\in\Sscr: \text{$S$ has type $\tau$ at $C$}} p_S$ for every type $\tau$ and $C\in C^{\xi}\setminus\{\delta(s),\delta(t)\}$.
We apply the previous lemma from left to right, and at each cut as long as possible.
In order to obtain a polynomial-time algorithm, we will round all $p_S$ down to integer multiples of $\frac{\epsilon}{n^2}$,
for some small positive constant $\epsilon$.

\begin{corollary}
\label{reassembleright}
For any constants $\frac{3}{2}<\xi<2$ and $\epsilon>0$
there is a polynomial-time algorithm which, given an instance $(V,s,t,c)$ and an optimum solution $x^*$ of {\rm (\ref{stpathlp})}, a 
set $\Sscr_+\subseteq\Sscr$ and numbers $p_S$ with $\frac{n^2}{\epsilon} p_S \in\mathbb{N}$ for $S\in\Sscr_+$
and $\sum_{S\in\Sscr_+}p_S\le 1$,
computes another 
set $\bar\Sscr_+\subseteq\Sscr$ and numbers $\bar p_S$ with $\frac{n^2}{\epsilon} \bar p_S \in\mathbb{N}$ for $S\in\bar\Sscr_+$,
such that $\sum_{S\in\Sscr_+}p_S\chi^S=\sum_{S\in\bar\Sscr_+}\bar p_S\chi^S$ and
\begin{equation}
\label{reassemblerightresult}
\min\{\barp{011},\barp{120}\} = 0 
\quad \text{ and } \quad 
\barp{110} \le \p{110} + \barpgood
\quad \text{ and } \quad 
\barp{021} \le \p{021} + \barpgood
\end{equation}
for all $C\in C^{\xi}\setminus\{\delta(s),\delta(t)\}$.
\end{corollary}

\prove
We first compute the $\xi$-narrow cuts (cf.\ Proposition \ref{computenarrowcuts}) and process them
from left to right, ignoring $\delta(s)$ and $\delta(t)$. Initially, $\bar\Sscr_+:=\Sscr_+$ and $\bar p := p$.
At each $C_i$ ($1\le i\le \ell'-1$) we check the types of all trees $S\in\bar\Sscr_+$.
Whenever we have two such trees $S_1$ and $S_2$ of type 120 and 011, respectively,
we set $\delta:= \min\{\bar p_{S_1},\bar p_{S_2}\}$, decrease $\bar p_{S_1}$ and $\bar p_{S_2}$ by $\delta$, and 
increase $\bar p_{S'_1}$ and $\bar p_{S'_2}$ by $\delta$, where $S_1'$ and $S'_2$ are chosen as in 
Lemma \ref{reassemblelemma}. 
If $S'_2$ but not $S_2$ has type 110 or 021 at $C_j$ for some $j>i$, then $S'_1$ has type \good{} at each of $C_{i+1},\ldots,C_j$, and
we maintain the properties
$\min\{\bar p^{\,C_h}_{\text{\rm 011}},\bar p^{\,C_h}_{\text{\rm 120}} \}=0$ for $h<i$ and
$\bar p^{\,C_j}_{\text{110}} \le p^{\,C_j}_{\text{110}} + \sum_{S\in\Sscr'} \bar p_S$
and
$\bar p^{\,C_j}_{\text{021}} \le p^{\,C_j}_{\text{021}} + \sum_{S\in\Sscr'} \bar p_S$
for all $j$,
where $\Sscr':=\{S\in\Sscr,\, S \text{ has type \good{} at $C_j$ and type 121 or \good{} at $C_i,\ldots,C_{j-1}$}\}$.

We remove a tree $S$ from $\bar S_+$ if $\bar p_S$ drops to zero.
Note that at any stage, all $\bar p_S$ (and $\delta$) are integer multiples of $\frac{\epsilon}{n^2}$, 
so there are never more than $\frac{n^2}{\epsilon}$ trees in $\bar S_+$.
\endproof

By symmetry, we also have:

\begin{corollary}
\label{reassembleleft}
For any constants $\frac{3}{2}<\xi<2$ and $\epsilon>0$
there is a polynomial-time algorithm which, given an instance $(V,s,t,c)$ and an optimum solution $x^*$ of {\rm (\ref{stpathlp})}, a 
set $\Sscr_+\subseteq\Sscr$ and numbers $p_S$ with $\frac{n^2}{\epsilon} p_S \in\mathbb{N}$ for $S\in\Sscr_+$
and $\sum_{S\in\Sscr_+}p_S\le 1$,
computes another 
set $\bar\Sscr_+\subseteq\Sscr$ and numbers $\bar p_S$ with $\frac{n^2}{\epsilon} \bar p_S \in\mathbb{N}$ for $S\in\bar\Sscr_+$,
such that $\sum_{S\in\Sscr_+}p_S\chi^S=\sum_{S\in\bar\Sscr_+}\bar p_S\chi^S$ and
\begin{equation}
\label{reassembleleftresult}
\min\{\barp{110},\barp{021}\} = 0 
\quad \text{ and } \quad 
\barp{011} \le \p{011} + \barpgood
\quad \text{ and } \quad 
\barp{120} \le \p{120} + \barpgood
\end{equation}
for all $C\in C^{\xi}\setminus\{\delta(s),\delta(t)\}$.
\endproof
\end{corollary}
 
We conclude:

\begin{theorem}
\label{reassemblethm}
For any constants $\frac{3}{2}<\xi<2$ and $\epsilon>0$
there is a polynomial-time algorithm which, given an instance $(V,s,t,c)$ and 
an optimum solution $x^*$ of {\rm (\ref{stpathlp})}, 
computes a set $\Sscr_+\subseteq \Sscr$ of trees such that there exists a 
distribution $p$ on $\Sscr$ with $x^*=\sum_{S\in\Sscr}p_S\chi^S$
and $p_S=0$ for all $S\in\Sscr\setminus\Sscr_+$
and
\begin{equation}
\label{reassembleresult}
\min\left\{\, \p{120}+\p{021},\ \p{011}+\p{110},\ \p{011}+\p{021},\ \p{120}+\p{110} \,\right\} 
\ \le \ \pgood + \epsilon
\end{equation}
for all $C\in C^{\xi}\setminus\{\delta(s),\delta(t)\}$.
\end{theorem}

\prove
First compute any distribution $p'$ with 
$x^*=\sum_{S\in\Sscr}p'_S\chi^S$ and with less than $n^2$ trees $S$ with $p'_S>0$, 
and the set $\Sscr'_+$ of these trees, using the ellipsoid method (\cite{GroLS81}) or the splitting-off technique (cf.\ \cite{GenW15}).
Then set $p''_S:= \frac{\epsilon}{n^2} \lfloor \frac{n^2}{\epsilon} p'_S \rfloor$ for all $S\in\Sscr$;
and let $\Sscr''_+$ be the set of trees with $p''_S>0$.
Note that $\sum_{S\in\Sscr} (p'_S-p''_S) < \epsilon$.

Then apply Corollary \ref{reassembleleft} to $\Sscr''_+$ and $p''$.
We get $\bar \Sscr_+$ and $\bar p$ with $\min\{\barp{110},\barp{021}\} = 0$ for all $C\in C^{\xi}\setminus\{\delta(s),\delta(t)\}$.
Applying Corollary \ref{reassembleright} to $\bar \Sscr_+$ and $\bar p$ yields $\bar{\bar \Sscr}_+$ and $\bar{\bar p}$
with $\min\{\dbarp{011},\dbarp{120}\} = 0 $ and $\min\{\dbarp{110},\dbarp{021}\} \le \dbarpgood$
for all $C\in C^{\xi}\setminus\{\delta(s),\delta(t)\}$.
Then $p:=\bar{\bar p} + p' - p''$ is a distribution as required.
We output $\Sscr_+:=\bar{\bar\Sscr}_+ \cup\Sscr'_+$.
\endproof

Note that our algorithm only needs the set $\Sscr_+$, which contains less than $\frac{n^2}{\epsilon}+n^2$ trees.
The above proof shows that we can also compute the distribution $p$ in polynomial time, but it is needed only for the analysis.

\section{Improved approximation ratio}
\label{sectionfirstapx}

We now show how to set the numbers $\gamma_{S,e}$ for $S\in\Sscr$ and $e\in I_S$
so that the total benefit (according to Definition \ref{defbenefit}) is large.

For some constant $\beta$ slightly larger than $\frac{2}{5}$ (to be determined later),
let $f(x):= \frac{\beta(2-x)(x-1)}{1-2\beta}$. 
By Lemma \ref{lemmaneedbenefit} and (\ref{boundeventreesincut}),
an average benefit of $f(x^*(C))$ from the trees will be sufficient for a narrow cut $C$.

Ideally, we would like to have $\gamma_{S,e}\ge f(x^*(C))$ if $e\in S\cap C$ and $|S\cap C|$ even,
and $1-\gamma_{S,e} \ge f(x^*(C))$ if $e\in S\cap C$ and $|S\cap C|=1$. 
But this may be impossible because $f(x^*(C))$ can be greater than $\frac{1}{2}$ for $\beta>\frac{2}{5}$.
Therefore we cut off at $\frac{1}{2}$. More precisely:

For $S\in\Sscr$ and $e\in I_S$ we define two numbers $0\le f_1,f_2\le\frac{1}{2}$.
If $|S\cap C|>1$ for all $C\in\Cscr$ with $e\in C$, then $f_1:=0$,
otherwise $f_1:=\min\{\frac{1}{2},\max\{f(x^*(C)): e\in C\in\Cscr, |S\cap C|=1\}\}$.
If $|S\cap C|$ is odd for all $C\in\Cscr$ with $e\in C$, then $f_2:=0$,
otherwise $f_2:=\min\{\frac{1}{2},\max\{f(x^*(C)): e\in C\in\Cscr, |S\cap C| \text{ even}\}\}$.
%
If $f_2<f_1$, then we set $\gamma_{S,e} := f_2$, otherwise we set $\gamma_{S,e}:=1-f_1$.

\subsection{Less critical cuts}

\begin{lemma}
\label{lesscriticalbenefit}
For all $C\in\Cscr$ with $f(x^*(C))\le\frac{1}{2}$ we have {\rm (\ref{enoughbenefit})}.
\end{lemma}

\prove
By the above choice of $\gamma_{S,e^S_C}$ ($S\in\Sscr$), we have $\gamma_{S,e^S_C} \ge f(x^*(C))$
if $|S\cap C|$ is even and $1-\gamma_{S,e^S_C} \ge f(x^*(C))$ if $|S\cap C|=1$.
Let $q:=\sum_{S\in\Sscr'}p_S$ where 
$\Sscr':=\bigl\{S\in\Sscr : |S\cap C| \text{ even},\, \frac{\beta(2-x^*(C))}{1-2\beta}<\gamma_{S,e^S_C}\bigr\}$.
We have (by Definition \ref{defbenefit})
\begin{eqnarray*}
\sum_{S\in\Sscr} p_S b_{S,C}
&\ge&  q \cdot \frac{\beta(2-x^*(C))}{1-2\beta} +
\left( \peven - q \right) \cdot f(x^*(C)) + 
\pone \cdot f(x^*(C)).
\end{eqnarray*}
As $\pone \ge 2-x^*(C) \ge
 \frac{2-x^*(C)}{x^*(C)-1} \, \peven \ge \frac{2-x^*(C)}{x^*(C)-1} \, (\peven-q)$
by (\ref{boundonetreesincut}) and (\ref{boundeventreesincut}), we get
$$
\sum_{S\in\Sscr} p_S b_{S,C}
\ \ge \ 
q \cdot \frac{\beta(2-x^*(C))}{1-2\beta}+
 f(x^*(C)) \left(1+ \frac{2-x^*(C)}{x^*(C)-1} \right) \left( \peven - q \right) \\
\ = \ 
\frac{\beta (2-x^*(C)) \peven}{1-2\beta}.
\vspace*{-2mm}
$$
\endproof

So we need to analyze only the remaining cuts (with $f(x^*(C))>\frac{1}{2}$ and hence $x^*(C)\approx 1.5$,
more precisely with
$\frac{3}{2}-\sqrt{\frac{5}{4}-\frac{1}{2\beta}} \le x^*(C) \le \frac{3}{2}+\sqrt{\frac{5}{4}-\frac{1}{2\beta}}$).

\subsection{Most critical cuts}

Now we consider the cuts $C\in\Cscr$ with $f(x^*(C)) > \frac{1}{2}$.
We first establish what we outlined in Subsection \ref{morebenefit}:

\begin{lemma}
\label{easybenefitbounds}
Let $0.4\le \beta < 0.5$ and $1.7 \le \xi\le 1.8$.
Let $C\in\Cscr^{\xi}\setminus\{\delta(s),\delta(t)\}$ with $f(x^*(C))\ge \frac{1}{2}$, and $S\in\Sscr$ with $|S\cap C|=1$ or $|S\cap C|$ even. 
Then:
\begin{itemize}
\item
$b_{S,C} \ge \frac{1}{2}$.
\item
Moreover, if $e^S_C\notin C_{\leftarrow} \cup C_{\rightarrow}$ or $S\cap C'\not=\{e^S_C\}$ for all $C'\in\Cscr^{\xi}$,
then $b_{S,C} \ge 1-f(\xi)$.
\end{itemize}
\end{lemma}

\prove
First let $|S\cap C|=1$. Then $b_{S,C} = 1-\gamma_{S,e^S_C}$.
For $S$ and $e=e_C^S$ we have $f_1=\frac{1}{2}\ge f_2$, so $\gamma_{S,e^S_C}= f_2\le \frac{1}{2}$.
Moreover, if $e^S_C\notin C_{\leftarrow} \cup C_{\rightarrow}$, then $f_2 \le f(\xi)$.

Now let $|S\cap C|$ be even. Then 
$b_{S,C}:=\min\left\{\frac{\beta (2-x^*(C)) }{1-2\beta}, \gamma_{S,e^S_C} \right\}$.

We first observe that $x^*(C)<1.7$ and $f(\xi)\ge f(1.8) \ge 0.32$ and hence
$\frac{\beta(2-x^*(C))}{1-2\beta} = \frac{f(x^*(C))}{x^*(C)-1} > \frac{f(x^*(C))}{0.7} \ge \frac{5}{7} > 1-f(\xi)$. 

Now consider the second term. We have $f_2=\frac{1}{2}\ge f_1$, so $\gamma_{S,e^S_C}=1-f_1\ge\frac{1}{2}$.
Moreover, 
if $S\cap C'\not=\{e^S_C\}$ for all $C'\in\Cscr^{\xi}$ (this holds in particular
if $|S\cap C|\not=1$ and $e^S_C\notin C_{\leftarrow} \cup C_{\rightarrow}$), then 
$f_1 \le f(\xi)$.
\endproof

Now we can bound the benefit for most critical cuts, of course using a distribution according to Theorem \ref{reassemblethm}.




\begin{lemma}
\label{caseanalysis}
Let $0.4\le\beta<0.5$  and $1.7 \le \xi \le 1.8$ such that $\nu:=1-f(\xi)>\frac{1}{2}$. Let $\epsilon>0$.
Let $p$ be a distribution on $\Sscr$ with $x^*=\sum_{S\in\Sscr}p_S\chi^S$ and {\rm (\ref{reassembleresult})}.
Let $C\in\Cscr^{\xi}\setminus\{\delta(s),\delta(t)\}$ with $f(x^*(C))>\frac{1}{2}$
and $x^*(C)\ge 2-\frac{\xi}{3}$.
Then
\begin{equation}
\label{benefitclaim}
2 \sum_{S\in\Sscr} p_S b_{S,C} \ \ge \ 
1 + \left( 5-\frac{3}{2}(x^*(C)+\xi)  -\epsilon\right) \left(\nu-\frac{1}{2} \right) - \left( 4\nu-1 \right) \pmany.
\end{equation}
\end{lemma}

\prove
As before, let $C_{\leftarrow}$ and $C_{\rightarrow}$ be the adjacent $\xi$-narrow cuts left and right.

Let 
$l_S=|S\cap C\cap C_{\leftarrow}|$, $m_S=|S\cap C|$, and $r_S=|S\cap C\cap C_{\rightarrow}|$.
Note that $\sum_{S\in\Sscr} p_S m_S = x^*(C)$ and, using Lemma \ref{smallintersection},
$\sum_{S\in\Sscr} p_S l_S  = x^*(C_{\leftarrow} \cap C) 
\le \frac{1}{2}(x^*(C_{\leftarrow})+x^*(C)) - 1 \le \frac{1}{2}(\xi + x^*(C)) - 1$.
Analogously, $\sum_{S\in\Sscr} p_S r_S \le \frac{1}{2}(\xi + x^*(C)) - 1$.

For all $S\in\Sscr$ with $m_S\ge 3$, we have
\begin{equation}
\label{benefitifmany}
2b_{S,C} - (m_S+1) (\nu-\textstyle\frac{1}{2}) +(4\nu-1) \lfloor\frac{m_S-1}{2}\rfloor 
\ \ge \ 1.
\end{equation}
(If $m_S$ is odd, then $b_{S,C}= 0$ and thus
$2b_{S,C} - (m_S+1) (\nu-\textstyle\frac{1}{2}) +(4\nu-1) \lfloor\frac{m_S-1}{2}\rfloor 
= m_S \nu -3 \nu + 1 \ge 1$.
If $m_S$ is even, then $b_{S,C}\ge \frac{1}{2}$ and thus
$2b_{S,C} - (m_S+1) (\nu-\textstyle\frac{1}{2}) +(4\nu-1) \lfloor\frac{m_S-1}{2}\rfloor 
\ge m_S \nu -5 \nu + \frac{5}{2} \ge \frac{3}{2}$.)

Now we distinguish four cases (cf.\ (\ref{reassembleresult})).

\newcommand{\nominus}{\phantom{$-$}}
\begin{table}[t]
\begin{center}
\renewcommand\arraystretch{1.15}
\begin{tabular}{|c||c||c|c|c||c|c|c|c|}\hline
                                            &     $b_{S,C}$             & $l$ & $m$ & $r$  & 
                                            \small Case 1 &  \small Case 2 & \small Case 3 & \small Case 4 \\\hline
                                            & benefit & \multicolumn{3}{c||}{number of edges in} & 
                                            \small $l+r-m$ & \small $l+r+m$ & \small \ $2l+r$ \ &  \small \ $l+2r$ \ \\
 type                                    &  $\ge$  & \small $C_{\leftarrow}\cap C$ & \small $C$ & \small $C \cap C_{\rightarrow}$ & 
                                            \small $\ge$ & \small $\ge$ & $\ge$ & $\ge$ \\\hline
 \textcolor{darkgreen}{010} & \textcolor{darkgreen}{$\nu$} & 0 & 1 & 0 & $-1$ & 1 & 0 & 0 \\
 \textcolor{orange}{011}      & $\frac{1}{2}$                          & 0 & 1 & 1 & \nominus 0 & \textcolor{orange}{2} &  \textcolor{orange}{1} & 2 \\
 \textcolor{purple}{110}       & $\frac{1}{2}$                          & 1 & 1 & 0 & \nominus 0 & \textcolor{purple}{2} & 2 & \textcolor{purple}{1} \\
 \textcolor{blue}{111}           & $\frac{1}{2}$                          & 1 & 1 & 1 & \nominus 1 & 3 & 3 & 3 \\\hline
 \textcolor{darkgreen}{020} & \textcolor{darkgreen}{$\nu$} & 0 & 2 & 0 & $-2$ & 2 & 0 & 0 \\
 \textcolor{purple}{021}       & $\frac{1}{2}$                          & 0 & 2 & 1 &  \textcolor{purple}{$-1$} & 3 & \textcolor{purple}{1} & 2 \\
 \textcolor{orange}{120}      & $\frac{1}{2}$                          & 1 & 2 & 0 &  \textcolor{orange}{$-1$} & 3 & 2 & \textcolor{orange}{1} \\
 \textcolor{blue}{022}          & $\frac{1}{2}$                          & 0 & 2 & 2 & \nominus 0 & 4 & 2 & 4 \\
 \textcolor{blue}{220}          & $\frac{1}{2}$                          & 2 & 2 & 0 & \nominus 0 & 4 & 4 & 2 \\
 \textcolor{blue}{121}          & $\frac{1}{2}$                          & 1 & 2 & 1 & \nominus 0 & 4 & 3 & 3 \\\hline
 \good                                 & $\nu$                 & $l$ & $2$ & $\ge 1-l$ & $-1$ & 2 & 1 & 1 \\
 \good                                 & $\frac{1}{2}$      & $l$ & $2$ & $\ge 3-l$ & \nominus 1 & 4 & 3 & 3 \\
 \good                                 & 0                        & $\!\!\!\!\!\ge 0$ & $m\ge 3$ & $\!\!\!\!\!\ge 0$         & $-m$ & $m$ & 0 & 0 \\  
\hline
\end{tabular}     
\end{center}
\caption{\label{benefittable}
The different types and their contributions.
The bounds on the benefit (second column) follow from Lemma \ref{easybenefitbounds}; here we use $f(x^*(C))\ge\frac{1}{2}$.
After reassembling, except for an $\epsilon$ fraction of the trees,
the two orange types (011 and 120) cannot occur simultaneously,
and if the two purple types (110 and 021) occur simultaneously, this is compensated by \good{} types.}
\end{table}

\bigskip\noindent {\bf Case 1:} 
$\p{120}+\p{021}\le \pgood + \epsilon$.
\smallskip

For $S\in\Sscr$, let $a_S=1$ if $S$ has type 120 or 021 at $C$,
$a_S=-1$ if $S$ has type \good{} at $C$,
and $a_S=0$ otherwise.
Note that $\sum_{S\in\Sscr}p_Sa_S \le \epsilon$.

Then for all trees $S\in\Sscr$ we have (cf.\ Lemma \ref{easybenefitbounds} and Table \ref{benefittable}):
\begin{itemize}
\item $m_S\ge 3$, or
\item $b_{S,C} \ge \frac{1}{2}$ and $l_S+r_S-m_S+a_S \ge 0$, or
\item $b_{S,C} \ge \nu$ and $l_S+r_S-m_S+a_S \ge -2$,
\end{itemize}
and hence (using (\ref{benefitifmany}) for the case $m_S\ge 3$)
$$2b_{S,C} + (l_S+r_S-m_S+a_S) (\nu-\textstyle\frac{1}{2}) +(4\nu-1) \lfloor\frac{m_S-1}{2}\rfloor 
\ \ge \ 1.$$

Taking the weighted sum,  this implies (using (\ref{boundmanytreesincut}))
\begin{eqnarray*}
2 \sum_{S\in\Sscr} p_S b_{S,C}   &\ge& 1 + \left(2-\xi-\epsilon \right) (\nu-\textstyle\frac{1}{2}) - (4\nu-1)\pmany.
\end{eqnarray*}
As $x^*(C)\ge 2-\frac{\xi}{3}$ implies $2-\xi \ge 5-\frac{3}{2}(x^*(C)+\xi)$, we obtain (\ref{benefitclaim}).

\bigskip\noindent {\bf Case 2:} 
$\p{011}+\p{110}\le \pgood + \epsilon$.
\smallskip

For $S\in\Sscr$, let $a_S=1$ if $S$ has type 011 or 110 at $C$,
$a_S=-1$ if $S$ has type \good{} at $C$,
and $a_S=0$ otherwise.
Note that $\sum_{S\in\Sscr}p_Sa_S \le \epsilon$.

Then for all trees $S\in\Sscr$ we have (cf.\ Lemma \ref{easybenefitbounds} and Table \ref{benefittable}):
\begin{itemize}
\item $m_S\ge 3$, or
\item $b_{S,C} \ge \frac{1}{2}$ and $l_S+r_S+m_S+a_S \ge 3$, or
\item $b_{S,C} \ge \nu$ and $l_S+r_S+m_S+a_S \ge 1$,
\end{itemize}
and hence (using again (\ref{benefitifmany}) for the case $m_S\ge 3$)
$$2b_{S,C} + (l_S+r_S+m_S + a_S - 3) (\nu-\textstyle\frac{1}{2}) + (4\nu-1) \lfloor\frac{m_S-1}{2}\rfloor 
\ \ge \ 1.$$
Taking the weighted sum, this implies
\begin{eqnarray*}
2 \sum_{S\in\Sscr} p_S b_{S,C}   &\ge& 1 + \left(5-2x^*(C)-\xi -\epsilon \right) (\nu-\textstyle\frac{1}{2}) -  (4\nu-1)\pmany,
\end{eqnarray*}
implying (\ref{benefitclaim}) because $\xi\ge x^*(C)$.

\bigskip\noindent {\bf Case 3:} 
$\p{011}+\p{021}\le \pgood+\epsilon$.
\smallskip

For $S\in\Sscr$, let $a_S=1$ if $S$ has type 011 or 021 at $C$,
$a_S=-1$ if $S$ has type \good{} at $C$,
and $a_S=0$ otherwise.
Note that $\sum_{S\in\Sscr}p_Sa_S \le \epsilon$.

Then for all trees  $S\in\Sscr$ we have (cf.\ Lemma \ref{easybenefitbounds} and Table \ref{benefittable}):
\begin{itemize}
\item $m_S\ge 3$, or
\item $b_{S,C} \ge \frac{1}{2}$ and $2l_S+r_S+a_S \ge 2$, or
\item $b_{S,C} \ge \nu$ and $2l_S+r_S+a_S \ge 0$,
\end{itemize}
and hence (using once more (\ref{benefitifmany}) for the case $m_S\ge 3$)
$$2b_{S,C} + (2l_S+r_S+a_S-2) (\nu-\textstyle\frac{1}{2}) + (4\nu-1) \lfloor\frac{m_S-1}{2}\rfloor 
\ \ge \ 1.$$
Taking the weighted sum, this implies
\begin{eqnarray*}
2 \sum_{S\in\Sscr} p_S b_{S,C}  
&\ge& 1 + \left(5-\textstyle\frac{3}{2}(x^*(C)+\xi) -\epsilon \right) \left(\nu-\textstyle\frac{1}{2} \right) - (4\nu-1) \pmany, 
\end{eqnarray*}
i.e., (\ref{benefitclaim}).

\bigskip\noindent {\bf Case 4:} 
$\p{110}+\p{120}\le \pgood+\epsilon$.
This is  symmetric to Case 3.

\bigskip
So (\ref{benefitclaim}) is proved in all cases.
\endproof

\subsection{Setting the constants}

We now obtain our main result easily:

\begin{theorem}
\label{mainthm}
If $\xi=1.73$ and $\epsilon=0.01$ and $p$ is a distribution as obtained in Theorem \ref{reassemblethm},
then $\bomc(p) \le (2-\beta) c(x^*)$ for $\beta=0.401$. In particular, we have an
$1.599$-approximation algorithm for the $s$-$t$-path TSP, and the integrality ratio of {\rm (\ref{stpathlp})} is at most
$1.599$.
\end{theorem}

\prove
We have $\nu= 1-f(\xi) >0.6$.
We use Lemma \ref{lemmaneedbenefit} and need to show (\ref{enoughbenefit}).

Let $C\in\Cscr$ be a narrow cut.  
If $f(x^*(C))\le\frac{1}{2}$, we have shown (\ref{enoughbenefit}) in Lemma \ref{lesscriticalbenefit}.
So let now $C\in\Cscr$ be a narrow cut with $f(x^*(C))>\frac{1}{2}$. 
Note that $1.44 < x^*(C) < 1.56$ and thus in particular
$C\in C^{\xi}\setminus\{\delta(s),\delta(t)\}$ and $x^*(C)\ge 2-\frac{\xi}{3}$.
We apply Lemma \ref{caseanalysis} to $C$.

The constants $\xi$, $\epsilon$, and $\beta$ are chosen so that
\begin{equation}
\label{boundonbeta}
1 + \left(5-\frac{3}{2}(x^*(C)+\xi) -\epsilon \right) \left(\nu-\frac{1}{2} \right)
\ \ge \
\frac{2\beta}{1-2\beta} \, (x^*(C)-1) (2-x^*(C))
\end{equation}
holds for all values of $x^*(C)$.

Moreover,
$\beta 
\ge
 \frac{3}{6+4\xi(2-\xi)}$,
so
$4\beta(2-\xi)
\ge
 3-6\beta-4\beta(2-\xi)(\xi-1)$
and hence $\frac{2\beta}{1-2\beta}(2-x^*(C)) \ge \frac{2\beta}{1-2\beta}(2-\xi) 
\ge \frac{3-6\beta-4\beta(2-\xi)(\xi-1)}{2(1-2\beta)} 
= \frac{3-4f(\xi)}{2} 
= \frac{4\nu -1}{2}$.
Therefore, using (\ref{boundmanytreesincut}) and (\ref{boundeventreesincut}),
\begin{eqnarray*}
&& \hspace*{-5cm} \frac{2\beta (2-x^*(C)) \peven}{1-2\beta}+ (4\nu-1)\pmany \\
&=& \peven \left(\frac{2\beta(2-x^*(C))}{1-2\beta} - \frac{4\nu-1}{2}\right) +  \frac{4\nu-1}{2}(x^*(C)-1) \hspace*{-4cm} \\
&\le& (x^*(C)-1)\left(\frac{2\beta(2-x^*(C))}{1-2\beta} - \frac{4\nu-1}{2}\right) +  \frac{4\nu-1}{2}(x^*(C)-1) \hspace*{-4cm} \\
&=& \frac{2\beta}{1-2\beta} (x^*(C)-1) (2-x^*(C)).
\end{eqnarray*}

Together with (\ref{benefitclaim}) and (\ref{boundonbeta}), this directly implies (\ref{enoughbenefit}).
\endproof


\section{Enhancements}

The constants $\xi$ and $\beta$ in the previous section are not optimal, but they are close.
The bounds are almost tight for Case 3 (and 4) and $x^*(C)\approx 1.52$ and $\pmany=0$.
However, we now suggest two ideas for a refined analysis that leads to a further improvement.

Firstly, since (in contrast to \citegenitivs{Seb13} analysis) the worst case does not occur in $x^*(C)=1.5$, but in a slightly
larger value, one can increase $\beta$ and hence improve the approximation ratio 
by increasing the $\gamma$-values slightly.

Secondly, the analysis in Case 3 (and 4) of Lemma \ref{caseanalysis} can be refined
(and the analysis was not tight in Case 1 and 2 anyway), as we will indicate now.
Consider a critical cut $C$ in Case 3, and assume for simplicity $\pmany=0$.
Let $\frac{3}{2}\le\xi'\le 2$ with $f(\xi')=\frac{1}{2}$, and let $C^{\xi'}_{\leftarrow}$
be the next $\xi'$-narrow cut to the left of $C$.
We have benefit
$\frac{1}{2}$ for at most a $\frac{1}{2}x(C\cap C^{\xi}_{\rightarrow}) + x(C\cap C^{\xi'}_{\leftarrow})
\le \frac{3}{4}x(C) + \frac{1}{4}\xi + \frac{1}{2} \xi' - \frac{3}{2}$ fraction of the trees,
and larger benefit for the others.
More generally, we have benefit at most
$1- f(y)$ for at most a $\frac{1}{2}x(C\cap C^{\xi}_{\rightarrow}) + x(C\cap C^y_{\leftarrow})
\le \frac{3}{4}x(C) + \frac{1}{4}\xi + \frac{1}{2}y - \frac{3}{2}$ fraction of the trees,
for all $y\in [\xi', \xi]$.
See Figure 5.

\begin{figure}
\begin{center}
\begin{tikzpicture}[thick, minimum size=0, inner sep=0, scale=0.9]
  \node[circle] (b2) at (-5,6.3) {};
  \node[circle] (b6) at (-5,4.7) {};
  \node[circle] (c3) at (-3,5.5) {};
  \node[circle] (c7) at (-3,4.0) {};
  \node[circle] (d1) at (-1,7) {};
  \node[circle] (d21) at (-1,6.2) {};
  \node[circle] (d31) at (-1,5.4) {};
  \node[circle] (d8) at (-1,3.3) {};
  \node[circle] (d81) at (-1,3.2) {};
  \node[circle] (d9) at (-1,2.5) {};
  \node[circle] (d91) at (-1,2.4) {};
  \node[circle] (e1) at (1,7) {};
  \node[circle] (e2) at (1,6.3) {};
  \node[circle] (e21) at (1,6.2) {};
  \node[circle] (e3) at (1,5.5) {};
  \node[circle] (e31) at (1,5.4) {};
  \node[circle] (e6) at (1,4.7) {};
  \node[circle] (e7) at (1,4.0) {};
  \node[circle] (e8) at (1,3.3) {};
  \node[circle] (e81) at (1,3.2) {};
 \node[circle] (f9) at (3,2.5) {};
  \node[circle] (f91) at (3,2.4) {};
\draw[darkgreen] (d1) -- (e1);
\draw[orange] (b2) -- (e2);
\draw[orange] (d21) -- (e21);
\draw[orange] (c3) -- (e3);
\draw[orange] (d31) -- (e31);
\draw[purple] (b6) -- (e6);
\draw[purple] (c7) -- (e7);
\draw[darkgreen] (d8) -- (e8);
\draw[darkgreen] (d81) -- (e81);
\draw[blue] (d9) --  (f9);
\draw[blue] (d91) -- (f91);
\draw[grey] (-6,2)--(-6,7.4);
\draw[grey] (-4,2)--(-4,7.4);
\draw[grey] (-2,2)--(-2,7.4);
\draw[grey] (0,2)--(0,7.4);
\draw[grey] (2,2)--(2,7.4);
\node[grey] at (-6,7.8) {$C_{\leftarrow}^{\xi'}$};
\node[grey] at (-4,7.8) {$C_{\leftarrow}^{y}$};
\node[grey] at (-2,7.8) {$C_{\leftarrow}^{\xi}$};
\node[grey] at (0,7.8) {$C$};
\node[grey] at (2,7.8) {$C_{\rightarrow}^{\xi}$};
\node at (4.5,7.8) {type};
\node at (-9,7.8) {benefit $\ge$};
 \node[darkgreen] at (4.5,7) {010};
 \node[orange] at (4.5,6.25) {120};
 \node[orange] at (4.5,5.45) {120};
 \node[purple]  at (4.5,4.7) {110};
 \node[purple]  at (4.5,4.0) {110};
 \node[darkgreen]  at (4.5,3.25) {020};
 \node[blue] at (4.5,2.45) {022};
 \node[darkgreen] at (-9,7) {\small $1-f(\xi)$};
\node[orange] at (-9,6.25) {\small $\frac{1}{2}$};
\node[orange] at (-9,5.45) {\small $1-f(y)$};
\node[purple] at (-9,4.7) {\small $\frac{1}{2}$};
\node[purple] at (-9,4.0) {\small $1-f(y)$};
\node[darkgreen] at (-9,3.25) {\small $1-f(\xi)$};
\node[blue] at (-9,2.45) {\small $\frac{1}{2}$};

\end{tikzpicture}
\end{center}
\caption{Refined analysis of Case 3.
\label{figcase3}}
\end{figure}
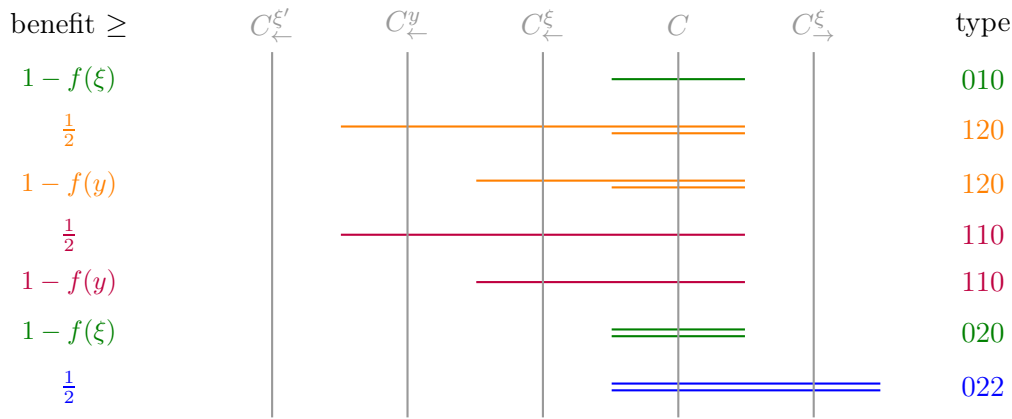

We have not performed the necessary calculations to obtain the best possible approximation ratio with these ideas.
However, it seems that the resulting improvements are rather small.
For a much better bound, we would probably need stronger reassembling results.

%
%
%
%
%

\section{Discussion \label{conclusion}}

Theorem \ref{mainthm} readily leads also to an improved approximation ratio for the
prize-collecting $s$-$t$-path TSP, simply by applying Theorem 6 of \cite{AnKS12}.
See \cite{GutXX00} for further applications.

Theorem \ref{mainthm} improves the best known upper bound on the integrality ratio
of the LP (\ref{stpathlp}). The best known lower bound is $\frac{3}{2}$, shown by the 
metric closure of an unweighted circuit. 
For metric closures of unweighted graphs the integrality ratio is indeed $\frac{3}{2}$
(even for $T$-tours), as \cite{SebV12} proved.
For general metrics, this remains open.

The most natural open question is of course to improve the approximation ratio further.
Our improvement was only small, but the reassembling technique could be more powerful than
we were able to prove. 
It seems that a stronger version of Lemma \ref{reassemblelemma} would be needed.
It would also be interesting to generalize our algorithm to the $T$-tour problem for general $T$
(\citegenitivs{Seb13} $\frac{8}{5}$-approximation algorithm and the previous algorithm of
\cite{CheFG15} work also for this more general problem).
%
Finally, applying our technique to other TSP variants would be very interesting.

\section*{Acknowledgement}
The author thanks the three referees for careful reading and excellent suggestions. 

\vspace*{5mm}

{\small
\newcommand{\bib}[3]{\bibitem[\protect\citeauthoryear{#1}{#2}]{#3}}

}

\newpage

\section*{Appendix: an Example}

The following example shows that we cannot get more benefit than \cite{Seb13}, 
no matter how we choose the numbers $\gamma_{S,e}$, 
if we represent the LP solution $x^*$ in an arbitrary way as
convex combination of spanning trees.

We show one instance with 20 vertices and 30 edges.
From this we obtain an infinite sequence of instances by extending the wall-like part in
the middle, inserting copies of the blue part, adding 4 vertices and 6 edges in each step.

The number next to an edge $e$ is $x^*(e)$; then $x^*$ is a feasible solution of (\ref{stpathlp}).
Grey vertical lines show the narrow cuts. We have $x^*(C)=\frac{3}{2}$ for all $C\in\Cscr\setminus\{\delta(s),\delta(t)\}$.

\begin{center}
\begin{tikzpicture}[thick, minimum size=6, inner sep=2, scale=1.02]
  \node at (-1.3,1.2) {$s$};
  \node[circle,draw] (s) at (-1,1.2) {};
  \node[circle,draw] (a1) at (0,2.4) {};
  \node[circle,draw] (a2) at (1,2.4) {};
  \node[circle,draw] (b) at (2,1.2) {};
  \node[circle,draw] (b2) at (2,0) {};
  \node[circle,draw] (c1) at (3,1.2) {};
  \node[circle,draw] (c2) at (3,2.4) {};
  \node[circle,draw] (c3) at (4,1.2) {};
  \node[circle,draw] (c4) at (4,2.4) {};
  \node[circle,draw,blue] (d) at (6,1.2) {};
  \node[circle,draw,blue] (d2) at (6,0) {};
  \node[circle,draw,blue] (e) at (8,1.2) {};
  \node[circle,draw,blue] (e2) at (8,2.4) {};
  \node[circle,draw] (f) at (10,1.2) {};
  \node[circle,draw] (f2) at (10,0) {};
  \node[circle,draw] (g) at (11,1.2) {};
  \node[circle,draw] (g2) at (11,2.4) {};
  \node[circle,draw] (h1) at (12,0) {};
  \node[circle,draw] (h2) at (13,0) {};
  \node[circle,draw] (t) at (14,1.2) {};
  \node at (14.3,1.2) {$t$};
\draw (a1) -- node[above] {\small $1$} (a2);
\draw (b) -- node[right] {\small $1$} (b2);
\draw (c1) -- node[right] {\small $1$} (c2);
\draw (c3) -- node[right] {\small $1$} (c4);
\draw[blue] (d) -- node[right] {\small $1$} (d2);
\draw[blue] (e) -- node[right] {\small $1$} (e2);
\draw (f) -- node[right] {\small $1$} (f2);
\draw (g) -- node[right] {\small $1$} (g2);
\draw (h1) -- node[below] {\small $1$} (h2);
\draw (a2) -- node[above] {\scriptsize $\frac{1}{2}$} (c2);
\draw (a1) -- node[below left] {\scriptsize $\frac{1}{4}\!$} (b2);
\draw (a2) -- node[below left] {\scriptsize $\frac{1}{2}\!$} (b);
\draw (b) -- node[above] {\scriptsize $\frac{1}{2}$} (c1);
\draw (c1) -- node[below] {\scriptsize $\frac{1}{2}$} (c3);
\draw (c2) -- node[above] {\scriptsize $\frac{1}{2}$} (c4);
\draw (c3) -- node[above] {\scriptsize $\frac{1}{2}$} (d);
\draw[blue] (d) -- node[above] {\scriptsize $\frac{1}{2}$} (e);
\draw[blue] (e) -- node[above] {\scriptsize $\frac{1}{2}$} (f);
\draw (f) -- node[above] {\scriptsize $\frac{1}{2}$} (g);
\draw (g) -- node[above right] {\scriptsize $\!\frac{1}{2}$} (h1);
\draw[blue] (d2) -- node[below] {\scriptsize $\frac{1}{2}$} (f2);
\draw (f2) -- node[below] {\scriptsize $\frac{1}{2}$} (h1);
\draw (g2) -- node[above right] {\scriptsize $\!\frac{1}{4}$} (h2);
\draw(s)[bend left] to node[above left] {\scriptsize $\frac{3}{4}$} (a1);
\draw[bend right] (s) to node[below left] {\scriptsize $\frac{1}{4}$} (b2);
\draw (c4) to node[above] {\scriptsize $\frac{1}{2}$} (e2);
\draw[blue] (e2) to node[above] {\scriptsize $\frac{1}{2}$} (g2);
\draw (b2) to node[below] {\scriptsize $\frac{1}{2}$} (d2);
\draw[bend left] (g2) to node[above right] {\scriptsize $\frac{1}{4}$} (t);
\draw[bend right] (h2) to node[below right] {\scriptsize $\frac{3}{4}$} (t);
\draw[lightgrey] (-0.35,-0.4)--(-0.35,2.8);
\draw[lightgrey] (0.65,-0.4)--(0.65,2.8);
\draw[lightgrey] (1.65,-0.4)--(1.65,2.8);
\draw[lightgrey] (2.65,-0.4)--(2.65,2.8);
\draw[lightgrey] (3.65,-0.4)--(3.65,2.8);
\draw[lightgrey] (5.15,-0.4)--(5.15,2.8);
\draw[lightgrey] (7.15,-0.4)--(7.15,2.8);
\draw[lightgrey] (9.15,-0.4)--(9.15,2.8);
\draw[lightgrey] (10.65,-0.4)--(10.65,2.8);
\draw[lightgrey] (11.5,-0.4)--(11.5,2.8);
\draw[lightgrey] (12.65,-0.4)--(12.65,2.8);
\draw[lightgrey] (13.65,-0.4)--(13.65,2.8);
\end{tikzpicture}
\end{center}

Consider the red sets in the following figure and all singletons; call this set of sets $\Uscr$.
Then the 30 vectors $\delta(U)$, $U\in\Uscr$, are easily shown to be linearly independent.
Since $x^*$ satisfies all constraints of (\ref{stpathlp}) that correspond to these sets with equality, 
 $x^*$ is indeed a vertex of the polytope defined by (\ref{stpathlp}). 
 Hence it is an optimum solution for some objective function.

\begin{center}
\begin{tikzpicture}[thick, minimum size=6, inner sep=2, scale=1.02]
  \node at (-1.3,1.2) {$s$};
  \node[circle,draw] (s) at (-1,1.2) {};
  \node[circle,draw] (a1) at (0,2.4) {};
  \node[circle,draw] (a2) at (1,2.4) {};
  \node[circle,draw] (b) at (2,1.2) {};
  \node[circle,draw] (b2) at (2,0) {};
  \node[circle,draw] (c1) at (3,1.2) {};
  \node[circle,draw] (c2) at (3,2.4) {};
  \node[circle,draw] (c3) at (4,1.2) {};
  \node[circle,draw] (c4) at (4,2.4) {};
  \node[circle,draw] (d) at (6,1.2) {};
  \node[circle,draw] (d2) at (6,0) {};
  \node[circle,draw] (e) at (8,1.2) {};
  \node[circle,draw] (e2) at (8,2.4) {};
  \node[circle,draw] (f) at (10,1.2) {};
  \node[circle,draw] (f2) at (10,0) {};
  \node[circle,draw] (g) at (11,1.2) {};
  \node[circle,draw] (g2) at (11,2.4) {};
  \node[circle,draw] (h1) at (12,0) {};
  \node[circle,draw] (h2) at (13,0) {};
  \node[circle,draw] (t) at (14,1.2) {};
  \node at (14.3,1.2) {$t$};
\draw (a1) -- node[above] {\small $1$} (a2);
\draw (b) -- node[right] {\small $1$} (b2);
\draw (c1) -- node[right] {\small $1$} (c2);
\draw (c3) -- node[right] {\small $1$} (c4);
\draw (d) -- node[right] {\small $1$} (d2);
\draw (e) -- node[right] {\small $1$} (e2);
\draw (f) -- node[right] {\small $1$} (f2);
\draw (g) -- node[right] {\small $1$} (g2);
\draw (h1) -- node[below] {\small $1$} (h2);
\draw (a2) -- node[above] {\scriptsize $\frac{1}{2}$} (c2);
\draw (a1) -- node[below left] {\scriptsize $\frac{1}{4}\!$} (b2);
\draw (a2) -- node[below left] {\scriptsize $\frac{1}{2}\!$} (b);
\draw (b) -- node[below] {\scriptsize $\frac{1}{2}$} (c1);
\draw (c1) -- node[below] {\scriptsize $\frac{1}{2}$} (c3);
\draw (c2) -- node[above] {\scriptsize $\frac{1}{2}$} (c4);
\draw (c3) -- node[above] {\scriptsize $\frac{1}{2}$} (d);
\draw (d) -- node[above] {\scriptsize $\frac{1}{2}$} (e);
\draw (e) -- node[above] {\scriptsize $\frac{1}{2}$} (f);
\draw (f) -- node[above] {\scriptsize $\frac{1}{2}$} (g);
\draw (g) -- node[above right] {\scriptsize $\!\frac{1}{2}$} (h1);
\draw (d2) -- node[below] {\scriptsize $\frac{1}{2}$} (f2);
\draw (f2) -- node[below] {\scriptsize $\frac{1}{2}$} (h1);
\draw (g2) -- node[above right] {\scriptsize $\!\frac{1}{4}$} (h2);
\draw(s)[bend left] to node[above left] {\scriptsize $\frac{3}{4}$} (a1);
\draw[bend right] (s) to node[below left] {\scriptsize $\frac{1}{4}$} (b2);
\draw (c4) to node[above] {\scriptsize $\frac{1}{2}$} (e2);
\draw (e2) to node[above] {\scriptsize $\frac{1}{2}$} (g2);
\draw (b2) to node[below] {\scriptsize $\frac{1}{2}$} (d2);
\draw[bend left] (g2) to node[above right] {\scriptsize $\frac{1}{4}$} (t);
\draw[bend right] (h2) to node[below right] {\scriptsize $\frac{3}{4}$} (t);
 \draw[red] (0.5,2.43) circle(0.85 and 0.35);
 \draw[red] (2,0.6) circle(0.35 and 0.85);
 \draw[red] (3,1.8) circle(0.35 and 0.85);
 \draw[red] (3.5,1.8) circle(1.1 and 1.1);
 \draw[red] (4,1.8) circle(0.35 and 0.85);
 \draw[red] (6,0.6) circle(0.35 and 0.85);
 \draw[red] (8,1.8) circle(0.35 and 0.85);
 \draw[red] (10,0.6) circle(0.35 and 0.85);
 \draw[red] (11,1.8) circle(0.35 and 0.85);
 \draw[red] (12.5,-0.03) circle(0.85 and 0.35);
\end{tikzpicture}
\end{center}

Now we will show a (bad) decomposition of $x^*$ into incidence vectors of trees. 
We have $x^*=\sum_{S\in\Sscr} p_S \chi^S$,
where $p_S=\frac{1}{4}$ for the following four trees and $p_S=0$ for all other spanning trees.
In the wall-like part (grey cuts 6--9 from left in the top figure), the four trees have types 
011, 110, 021, and 120. For each of these four trees $S$, each edge $e$ of its $s$-$t$-path $I_S$ within the wall-like part belongs
to one narrow cut $C$ with $|C\cap S|=1$ and one narrow cut $C'$ with $|C'\cap S|=2$.
No matter how we choose $\gamma_{S,e}$, the total benefit of this edge to both cuts is 1, so the average benefit that a 
narrow cut in the wall-like part receives is at most $\frac{1}{2}$.
This shows that reassembling trees is necessary.

We were unable to prove that the best-of-many Christofides algorithm
has no better approximation ratio than 1.6, but to obtain a better ratio a completely different analysis would be necessary. 
In fact, no better lower bound than 1.5 is known.

\begin{center}
\begin{tikzpicture}[thick, minimum size=6, inner sep=2, scale=1.02]
  \node at (-1.3,1.2) {$s$};
  \node[circle,draw] (s) at (-1,1.2) {};
  \node[circle,draw] (a1) at (0,2.4) {};
  \node[circle,draw] (a2) at (1,2.4) {};
  \node[circle,draw] (b) at (2,1.2) {};
  \node[circle,draw] (b2) at (2,0) {};
  \node[circle,draw] (c1) at (3,1.2) {};
  \node[circle,draw] (c2) at (3,2.4) {};
  \node[circle,draw] (c3) at (4,1.2) {};
  \node[circle,draw] (c4) at (4,2.4) {};
  \node[circle,draw] (d) at (6,1.2) {};
  \node[circle,draw] (d2) at (6,0) {};
  \node[circle,draw] (e) at (8,1.2) {};
  \node[circle,draw] (e2) at (8,2.4) {};
  \node[circle,draw] (f) at (10,1.2) {};
  \node[circle,draw] (f2) at (10,0) {};
  \node[circle,draw] (g) at (11,1.2) {};
  \node[circle,draw] (g2) at (11,2.4) {};
  \node[circle,draw] (h1) at (12,0) {};
  \node[circle,draw] (h2) at (13,0) {};
  \node[circle,draw] (t) at (14,1.2) {};
  \node at (14.3,1.2) {$t$};
\draw[turq] (a1) --  (a2);
\draw[red] (b) --  (b2);
\draw[red] (c1) --  (c2);
\draw[red] (c3) --  (c4);
\draw[red] (d) -- (d2);
\draw[red] (e) -- (e2);
\draw[red] (f) --  (f2);
\draw[red] (g) --  (g2);
\draw[red] (h1) --  (h2);
\draw[turq] (a2) -- (c2);
\draw[red] (a2) --  (b);
\draw[turq] (c2) -- (c4);
\draw[red] (c3) -- (d);
\draw[red] (e) -- (f);
\draw[red] (g) -- (h1);
\draw[turq] (s) [bend left] to (a1);
\draw[turq] (c4) to (e2);
\draw[turq] (e2) to  (g2);
\draw[turq,bend left] (g2) to (t);
\end{tikzpicture}
\end{center}

\begin{center}
\begin{tikzpicture}[thick, minimum size=6, inner sep=2, scale=1.02]
  \node at (-1.3,1.2) {$s$};
  \node[circle,draw] (s) at (-1,1.2) {};
  \node[circle,draw] (a1) at (0,2.4) {};
  \node[circle,draw] (a2) at (1,2.4) {};
  \node[circle,draw] (b) at (2,1.2) {};
  \node[circle,draw] (b2) at (2,0) {};
  \node[circle,draw] (c1) at (3,1.2) {};
  \node[circle,draw] (c2) at (3,2.4) {};
  \node[circle,draw] (c3) at (4,1.2) {};
  \node[circle,draw] (c4) at (4,2.4) {};
  \node[circle,draw] (d) at (6,1.2) {};
  \node[circle,draw] (d2) at (6,0) {};
  \node[circle,draw] (e) at (8,1.2) {};
  \node[circle,draw] (e2) at (8,2.4) {};
  \node[circle,draw] (f) at (10,1.2) {};
  \node[circle,draw] (f2) at (10,0) {};
  \node[circle,draw] (g) at (11,1.2) {};
  \node[circle,draw] (g2) at (11,2.4) {};
  \node[circle,draw] (h1) at (12,0) {};
  \node[circle,draw] (h2) at (13,0) {};
  \node[circle,draw] (t) at (14,1.2) {};
  \node at (14.3,1.2) {$t$};
\draw[turq] (a1) --  (a2);
\draw[red] (b) --  (b2);
\draw[red] (c1) --  (c2);
\draw[red] (c3) --  (c4);
\draw[red] (d) -- (d2);
\draw[red] (e) -- (e2);
\draw[red] (f) --  (f2);
\draw[red] (g) --  (g2);
\draw[red] (h1) --  (h2);
\draw[turq] (a2) -- (c2);
\draw[red] (b) -- (c1);
\draw[turq] (c2) -- (c4);
\draw[red] (d) -- (e);
\draw[red] (f) --  (g);
\draw[turq] (g2) -- (h2);
\draw[turq] (s) [bend left] to (a1);
\draw[turq] (c4) to (e2);
\draw[turq] (e2) to  (g2);
\draw[turq,bend right] (h2) to (t);
\end{tikzpicture}
\end{center}

\begin{center}
\begin{tikzpicture}[thick, minimum size=6, inner sep=2, scale=1.02]
  \node at (-1.3,1.2) {$s$};
  \node[circle,draw] (s) at (-1,1.2) {};
  \node[circle,draw] (a1) at (0,2.4) {};
  \node[circle,draw] (a2) at (1,2.4) {};
  \node[circle,draw] (b) at (2,1.2) {};
  \node[circle,draw] (b2) at (2,0) {};
  \node[circle,draw] (c1) at (3,1.2) {};
  \node[circle,draw] (c2) at (3,2.4) {};
  \node[circle,draw] (c3) at (4,1.2) {};
  \node[circle,draw] (c4) at (4,2.4) {};
  \node[circle,draw] (d) at (6,1.2) {};
  \node[circle,draw] (d2) at (6,0) {};
  \node[circle,draw] (e) at (8,1.2) {};
  \node[circle,draw] (e2) at (8,2.4) {};
  \node[circle,draw] (f) at (10,1.2) {};
  \node[circle,draw] (f2) at (10,0) {};
  \node[circle,draw] (g) at (11,1.2) {};
  \node[circle,draw] (g2) at (11,2.4) {};
  \node[circle,draw] (h1) at (12,0) {};
  \node[circle,draw] (h2) at (13,0) {};
  \node[circle,draw] (t) at (14,1.2) {};
  \node at (14.3,1.2) {$t$};
\draw[red] (a1) --  (a2);
\draw[red] (b) --  (b2);
\draw[red] (c1) --  (c2);
\draw[red] (c3) --  (c4);
\draw[red] (d) -- (d2);
\draw[red] (e) -- (e2);
\draw[red] (f) --  (f2);
\draw[red] (g) --  (g2);
\draw[turq] (h1) --  (h2);
\draw[red] (a2) --  (b);
\draw[red] (c1) -- (c3);
\draw[red] (c3) -- (d);
\draw[red] (e) -- (f);
\draw[red] (g) -- (h1);
\draw[turq] (d2) -- (f2);
\draw[turq] (f2) --  (h1);
\draw[turq,bend right] (s) to  (b2);
\draw[turq] (b2) to (d2);
\draw[turq,bend right] (h2) to (t);
\end{tikzpicture}
\end{center}

\begin{center}
\begin{tikzpicture}[thick, minimum size=6, inner sep=2, scale=1.02]
  \node at (-1.3,1.2) {$s$};
  \node[circle,draw] (s) at (-1,1.2) {};
  \node[circle,draw] (a1) at (0,2.4) {};
  \node[circle,draw] (a2) at (1,2.4) {};
  \node[circle,draw] (b) at (2,1.2) {};
  \node[circle,draw] (b2) at (2,0) {};
  \node[circle,draw] (c1) at (3,1.2) {};
  \node[circle,draw] (c2) at (3,2.4) {};
  \node[circle,draw] (c3) at (4,1.2) {};
  \node[circle,draw] (c4) at (4,2.4) {};
  \node[circle,draw] (d) at (6,1.2) {};
  \node[circle,draw] (d2) at (6,0) {};
  \node[circle,draw] (e) at (8,1.2) {};
  \node[circle,draw] (e2) at (8,2.4) {};
  \node[circle,draw] (f) at (10,1.2) {};
  \node[circle,draw] (f2) at (10,0) {};
  \node[circle,draw] (g) at (11,1.2) {};
  \node[circle,draw] (g2) at (11,2.4) {};
  \node[circle,draw] (h1) at (12,0) {};
  \node[circle,draw] (h2) at (13,0) {};
  \node[circle,draw] (t) at (14,1.2) {};
  \node at (14.3,1.2) {$t$};
\draw[red] (a1) --  (a2);
\draw[red] (b) --  (b2);
\draw[red] (c1) --  (c2);
\draw[red] (c3) --  (c4);
\draw[red] (d) -- (d2);
\draw[red] (e) -- (e2);
\draw[red] (f) --  (f2);
\draw[red] (g) --  (g2);
\draw[turq] (h1) --  (h2);
\draw[turq] (a1) --  (b2);
\draw[red] (b) -- (c1);
\draw[red] (c1) -- (c3);
\draw[red] (d) -- (e);
\draw[red] (f) --  (g);
\draw[turq] (d2) -- (f2);
\draw[turq] (f2) --  (h1);
\draw[turq] (s) [bend left] to (a1);
\draw[turq] (b2) to (d2);
\draw[turq,bend right] (h2) to (t);
\end{tikzpicture}
\end{center}

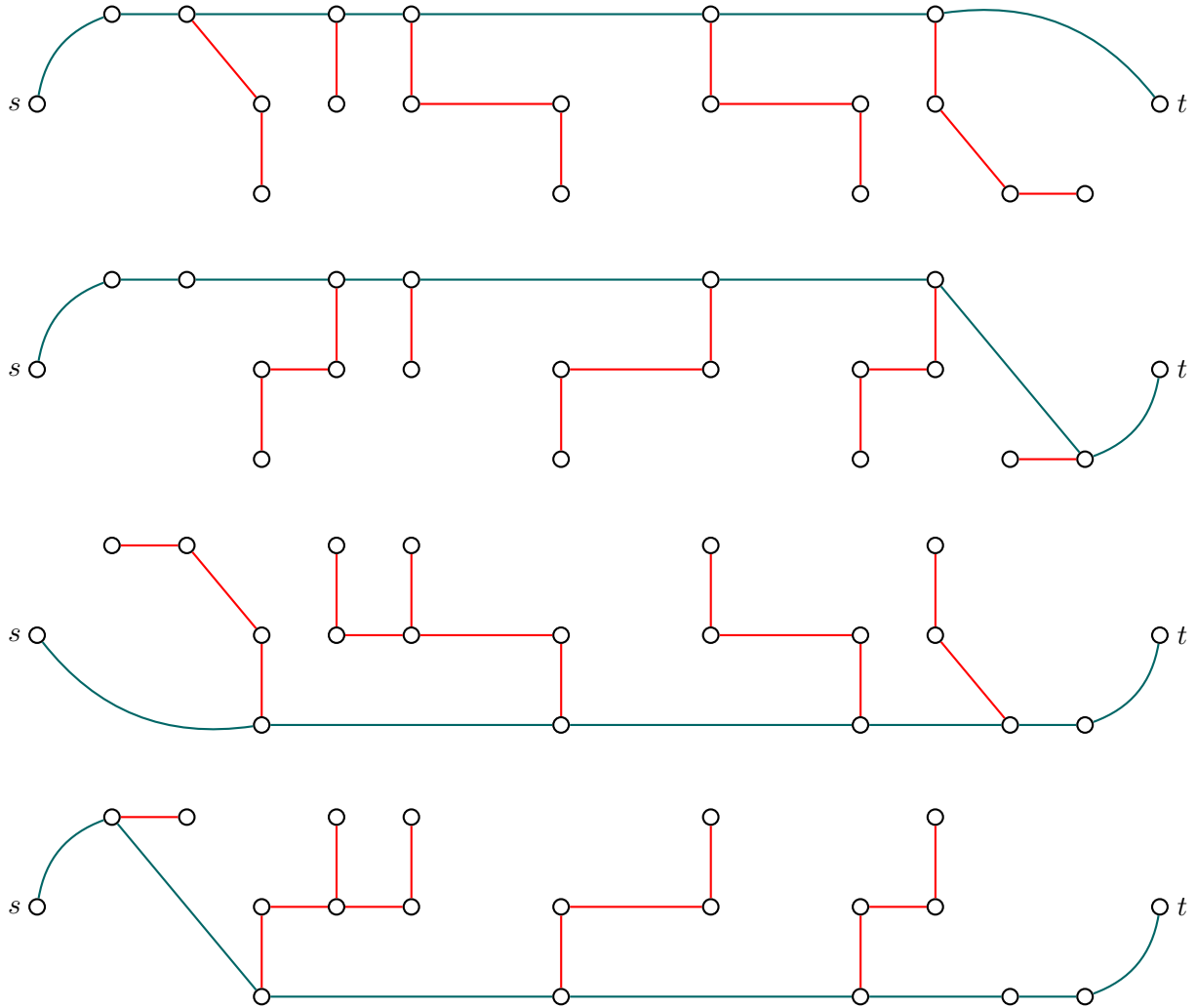
\end{document}